\documentclass[trackchanges,twocolumn]{aastex701}
\usepackage{amsmath}

\newcommand{\ha}{\mathrm{H\alpha}}
\newcommand{\hb}{\mathrm{H\beta}}

\newcommand{\lbha}{L_\mathrm{H\alpha,broad}}
\newcommand{\dvabs}{\Delta v_\mathrm{abs}}

\begin{document}

\shorttitle{
Demographics of Balmer Line Absorption in LRDs
%Balmer Absorption in Little Red Dots
}
%\shortauthor{Yanagisawa et al.}

\title{
ATLAS. II. \\
Extremely High Incidence of Balmer Line Absorption with Predominant Blueshifts in LRDs:\\
Statistical Insights through Comparison with Type 1 AGNs

%Extremely High Incidence of Balmer Line Absorbers in LRDs:\\
%Statistical Insights through Comparison with Type 1 AGNs

%Demographics of Balmer Line Absorption in Little Red Dots: \\
%High Incidence of Dense Gas Absorbers Compared to Low-$z$ Type 1 AGNs
}

\author[orcid=0009-0006-6763-4245]{Hiroto Yanagisawa}
\affiliation{Institute for Cosmic Ray Research, The University of Tokyo, 5-1-5 Kashiwanoha, Kashiwa, Chiba 277-8582, Japan}
\affiliation{Department of Physics, Graduate School of Science, The University of Tokyo, 7-3-1 Hongo, Bunkyo, Tokyo 113-0033, Japan}
\email[show]{yana@icrr.u-tokyo.ac.jp}

\author[0000-0002-1049-6658]{Masami Ouchi}
\affiliation{National Astronomical Observatory of Japan, 2-21-1 Osawa, Mitaka, Tokyo 181-8588, Japan}
\affiliation{Institute for Cosmic Ray Research, The University of Tokyo, 5-1-5 Kashiwanoha, Kashiwa, Chiba 277-8582, Japan}
\affiliation{Astronomical Science Program, Graduate Institute for Advanced Studies, SOKENDAI, 2-21-1 Osawa, Mitaka, Tokyo 181-8588, Japan}
\affiliation{Kavli Institute for the Physics and Mathematics of the Universe (WPI), The University of Tokyo, 5-1-5 Kashiwanoha, Kashiwa, Chiba 277-8583, Japan}
\email[]{ouchims@icrr.u-tokyo.ac.jp}

\author[orcid=0009-0004-4332-9225]{Tomokazu Kiyota}
\affiliation{National Astronomical Observatory of Japan, 2-21-1 Osawa, Mitaka, Tokyo 181-8588, Japan}
\affiliation{Astronomical Science Program, Graduate Institute for Advanced Studies, SOKENDAI, 2-21-1 Osawa, Mitaka, Tokyo 181-8588, Japan}
\email[]{tomokazu.kiyota@grad.nao.ac.jp}

\author[0000-0002-4225-4477]{Makoto Ando}
\affiliation{Institute for Cosmic Ray Research, The University of Tokyo, 5-1-5 Kashiwanoha, Kashiwa, Chiba 277-8582, Japan}
\email[]{mando@icrr.u-tokyo.ac.jp}

\author[0000-0002-6047-430X]{Yuichi Harikane}
\affiliation{Institute for Cosmic Ray Research, The University of Tokyo, 5-1-5 Kashiwanoha, Kashiwa, Chiba 277-8582, Japan}
\email[]{hari@icrr.u-tokyo.ac.jp}

\author[0009-0004-0381-7216]{Yuta Kageura}
\affiliation{Institute for Cosmic Ray Research, The University of Tokyo, 5-1-5 Kashiwanoha, Kashiwa, Chiba 277-8582, Japan}
\affiliation{Department of Physics, Graduate School of Science, The University of Tokyo, 7-3-1 Hongo, Bunkyo, Tokyo 113-0033, Japan}
\email[]{kageura@icrr.u-tokyo.ac.jp}

\author[0009-0000-1999-5472]{Minami Nakane}
\affiliation{Institute for Cosmic Ray Research, The University of Tokyo,
5-1-5 Kashiwanoha, Kashiwa, Chiba 277-8582, Japan}
\affiliation{Department of Physics, Graduate School of Science, The
University of Tokyo, 7-3-1 Hongo, Bunkyo, Tokyo 113-0033, Japan}
\email[]{nakanem@icrr.u-tokyo.ac.jp}  

\author[0000-0001-9011-7605]{Yoshiaki Ono}
\affiliation{Institute for Cosmic Ray Research, The University of Tokyo, 5-1-5 Kashiwanoha, Kashiwa, Chiba 277-8582, Japan}
\email[]{ono@icrr.u-tokyo.ac.jp}

\author[orcid=0009-0005-2897-002X]{Yui Takeda}
\affiliation{National Astronomical Observatory of Japan, 2-21-1 Osawa, Mitaka, Tokyo 181-8588, Japan}
\email[]{yui.takeda@grad.nao.ac.jp}
\affiliation{Astronomical Science Program, Graduate Institute for Advanced Studies, SOKENDAI, 2-21-1 Osawa, Mitaka, Tokyo 181-8588, Japan}

\begin{abstract}
% 250 words
We present the statistical properties of H$\alpha$ and H$\beta$ line absorption in little red dots (LRDs) at $z\simeq2.5$--7.2 using archival JWST/NIRSpec spectra from the DAWN JWST Archive and complementary NIRSpec/IFU observations. Among 40 LRDs with broad H$\alpha$ and [O~{\sc iii}] obtained with medium- or high-resolution gratings, 14 objects exhibit H$\alpha$ absorption. We find that the incidence of Balmer line absorption is $\sim35$\% ($=14/40$), significantly higher than that in SDSS low-$z$ type 1 AGNs ($\sim0.04$\%), demonstrating that Balmer line absorption occurs approximately 850 times more frequently in LRDs than in type 1 AGNs. We combine our 14 detections with 32 additional LRD Balmer absorbers from the literature, yielding a census of 46 absorbers. Their velocities span $\dvabs(\ha)\sim-430$ to $+140\ {\rm km\,s^{-1}}$, markedly narrower than the $-800$ to $+1600\ {\rm km\,s^{-1}}$ range of Balmer absorption in SDSS type~1 AGNs, for which our simulations confirm that the velocity difference is too large to be explained by detection incompleteness. The lower absolute absorber velocities in LRDs may partly reflect the shallower gravitational potential at their characteristic BLR radii. We also find that 38 of the 46 absorbers (83\%) are blueshifted, with only eight redshifted, indicating that most of Balmer absorbers are moving outward. An analytic model with radiation pressure suggests that most absorbers with $N_{\rm H}\gtrsim10^{24}\ {\rm cm^{-2}}$ remains gravitationally bound. The smaller number of redshifted (i.e., infalling) absorbers may indicate that outbound absorbers lose density: some return to the BLR, whereas others undergo stronger radiative acceleration and escape.

\end{abstract}


\section{Introduction}
% Strong Balmer break -> existence of dense, optically thick gas surrounding the central engine?
% this work
The advent of the James Webb Space Telescope (JWST) has revealed a numerous population of compact sources at high redshift with blue rest-frame ultraviolet (UV) continua and steeply rising red rest-frame optical continua, commonly referred to as little red dots (LRDs; \citealt{Kocevski+2023, Harikane+2023, Labbe+2025, Matthee+2024, Greene+2024, Kokorev+2024, Kocevski+2025, Akins+2025}). Spectroscopic observations have revealed broad Balmer emission lines in a large fraction of LRDs, providing compelling evidence for accreting black holes in at least a substantial subset of the population \citep{Greene+2024, Matthee+2024, Kocevski+2025, Hviding+2025}. Nevertheless, their multiwavelength properties differ markedly from those of conventional broad-line active galactic nuclei (AGNs). LRDs are typically weak or undetected in deep X-ray and radio observations \citep[e.g.,][]{Ananna+2024, Yue+2024}, exhibit little photometric variability \citep{Kokubo_Harikane_2025, Tee+2025, Burke+2025, ZZhang+2025a}, and often lack the prominent hot-dust emission characteristic of standard AGN spectral energy distributions \citep{Setton+2025, Casey+2025}, although recent radio and mid-infrared studies have revealed substantial diversity and evidence for AGN-heated dust in at least some of the population \citep{Gloudemans+2025, Barro+2026, Delvecchio+2025}. The physical origin of their characteristic V-shaped UV-to-optical spectral energy distributions and the structure of their nuclear environments therefore remain under active debate.

A particularly important clue to the nature of LRDs is the prominent spectral turnover around the Balmer limit. Strong continuum inflections in LRDs have been found to occur preferentially at the Balmer limit, suggesting that hydrogen opacity may play a central role in shaping their continua \citep{Setton+2025b}. While such features can arise from stellar populations with intrinsic Balmer breaks, several extreme LRDs exhibit Balmer-limit features that are difficult to reconcile with conventional stellar population models or with constraints on their stellar content \citep{Ji+2025, deGraaff+2025b, Naidu+2025}. These observations have motivated scenarios in which the central engine is surrounded by dense and optically thick gas. In particular, neutral hydrogen at densities of ($n_{\rm H}\sim10^{9}-10^{11}\ {\rm cm}^{-3}$) can develop a substantial population in the $n=2$ state and absorb the incident continuum blueward of the Balmer limit, producing a non-stellar Balmer break \citep{Inayoshi_Maiolino_2025, Ji+2025, deGraaff+2025b}. Related models involving optically thick gas envelopes or super-Eddington atmospheres can likewise reproduce key features of the red optical continua and Balmer-limit structure of LRDs, although the detailed origin and geometry of the continuum-emitting and absorbing gas remain uncertain \citep{Liu+2025, Naidu+2025, Torralba+2026}.

Balmer-line absorption provides a complementary and uniquely kinematic probe of the dense gas potentially associated with the unusual Balmer-limit features of LRDs. If the Balmer break is produced by hydrogen absorption, both the continuum break and the Balmer absorption lines arise from a substantial population of hydrogen in the $n=2$ state through bound-free and bound-bound opacity, respectively \citep{Inayoshi_Maiolino_2025}. However, whereas the physical origin of the continuum break can be degenerate, Balmer-line absorption directly reveals excited hydrogen in absorbing gas and, crucially, retains information on its line-of-sight velocity. The velocity offsets of the absorption features therefore provide a direct means of distinguishing outflowing, inflowing, and nearly systemic dense gas.

Balmer absorption is exceptionally rare among low-redshift type 1 AGNs \citep[e.g.,][]{Aoki+2006, Hall_2007, Ji+2012, Schulze+2018, Shangguan+2026}, yet H$\alpha$ and/or H$\beta$ absorption has now been reported in a growing number of LRDs and JWST-selected broad-line AGNs \citep{Matthee+2024, Kocevski+2025, Labbe+2024, Taylor+2025, Ji+2025, deGraaff+2025b, Naidu+2025, DEugenio+2026, LambridesAbs+2025, Torralba+2026}. The absorption is frequently blueshifted and can form P-Cygni-like profiles, motivating interpretations involving dense nuclear outflows \citep[e.g.,][]{Matthee+2026}. However, redshifted and nearly systemic absorbers have also been identified, indicating that Balmer absorption traces a broader range of circumnuclear gas motions, potentially including inflow and relatively slow-moving gas \citep{Matthee+2026, DEugenio+2026, Davis+2026}.

Recent sample-based studies have begun to investigate the incidence and physical properties of Balmer absorption and its connections to Balmer-break strength, continuum color, and gas covering factor \citep{Matthee+2026, Davis+2026, Chen+2026}. However, existing samples remain relatively small and heterogeneous, and the reported incidence of Balmer absorption varies substantially with spectral resolution and data quality. In particular, absorption features that are unresolved or diluted in low-resolution spectra can become apparent in medium- and high-resolution observations \citep{Davis+2026}.
Most recently, \citet{Juodzbalis+2026} presented a NIRSpec census of Balmer absorption in 47 type 1 AGNs at $z=2$--7 drawn from deep JWST surveys. However, Balmer absorption is rare, and each study has relied on a sample defined by its own selection and observing strategy, so the known absorbers remain few in number and fragmented across studies. This has limited the statistical significance of population-level inferences. 

This paper is the second in the Archival and Theoretical study of LRDs with AGN comparison across Surveys (ATLAS) series, in which we aim to provide physical interpretations of the LRD population through systematic comparisons with known AGN populations. In the first paper of the series \citep{Yanagisawa+2026}, we constructed a sample of compact broad-H$\alpha$ sources from archival JWST/NIRSpec spectra and showed that LRDs exhibit enhanced broad H$\alpha$ emission relative to low-$z$ type 1 AGNs at fixed continuum luminosity. Forthcoming papers in the series will extend these comparisons to other observational properties of LRDs.
In this work, we explore the statistical properties of Balmer line absorption in LRDs using archival JWST data. To improve the statistical significance of the analysis, we search the entire DJA for Balmer-absorption LRDs to assemble a large sample, and we further incorporate inclusively the absorbers reported in previous LRD studies \citep{Matthee+2026, Davis+2026, Chen+2026, Juodzbalis+2026}. We then place these absorbers in context by comparing them against the Balmer absorbers of low-$z$ type 1 AGNs and BAL quasars. 
%, and by interpreting all of them within a single physical framework that assesses, from the observed velocity together with the black hole mass and radiation field, whether the absorbing gas is gravitationally bound or escaping. Because the literature measurements remain heterogeneous in selection and methodology, this framework provides a common physical interpretation rather than a homogenization of the measurements. 
In Section~\ref{sec:data}, we summarize the parent sample and the working sample used for the absorption search. In Section~\ref{sec:line_profile_fitting}, we describe a method of the line-profile fitting. Section~\ref{sec:results} presents the incidence, velocity distribution, and equivalent widths of the Balmer absorbers. Section~\ref{sec:discussion} discusses the implications for the geometry and kinematics of dense gas in LRDs. We summarize our main conclusions in Section~\ref{sec:summary}.
Throughout this paper, we assume cosmology parameters based on the TT, TE, EE + lowE + lensing + BAO result from \cite{Planck+2020} with $H_0=67.66 \, \mathrm{km\,s^{-1}\,Mpc^{-1}}$, $\Omega_\mathrm{m}=0.30966$, and $\Omega_\mathrm{b}=0.04897$. %All magnitudes are in the AB system \citep{Oke_Gunn_1983}.

\begin{figure*}
    \centering
    \includegraphics[width=1\linewidth]{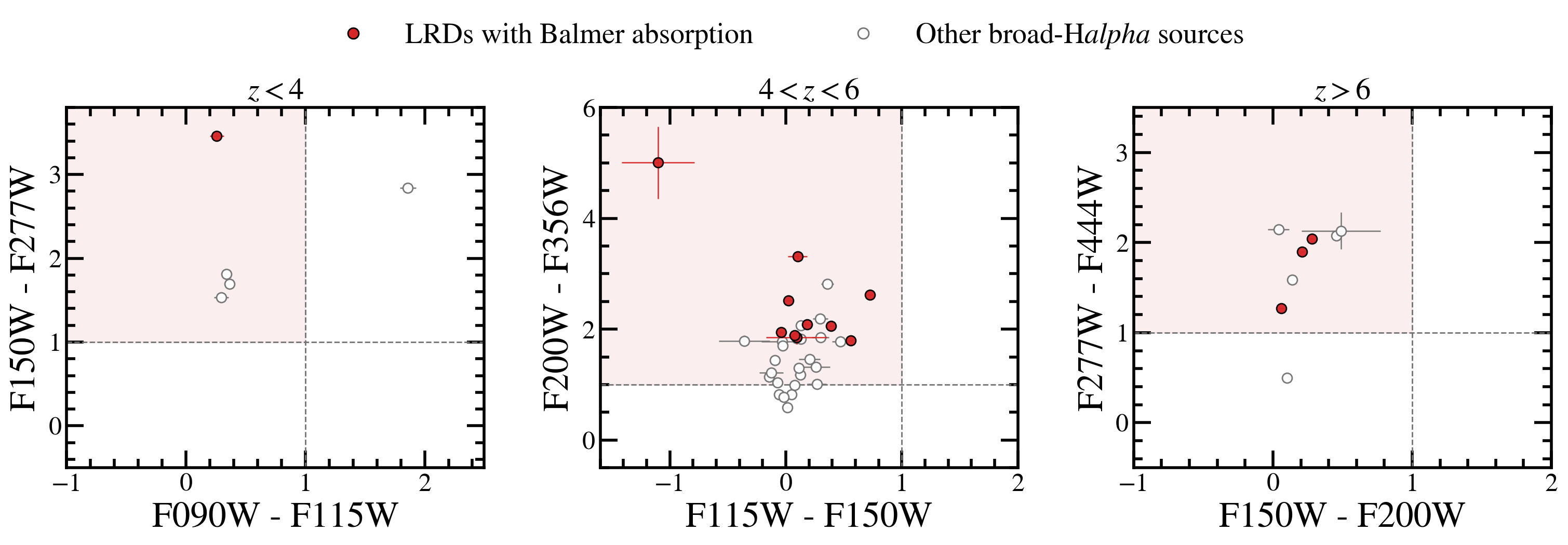}
    \caption{LRD color criteria presented in Equation~\eqref{eq:color}. From left to right, the panels show the criteria for $z<4$, $4<z<6$, and $z>6$. The pink shaded regions indicate the adopted color-selection windows. Filled red circles show adopted LRDs with H$\alpha$ absorption. Open gray circles show the remaining broad-H$\alpha$ sources in the parent sample. }
    \label{fig:color}
\end{figure*}

\begin{figure*}
    \centering
    \includegraphics[width=1\linewidth]{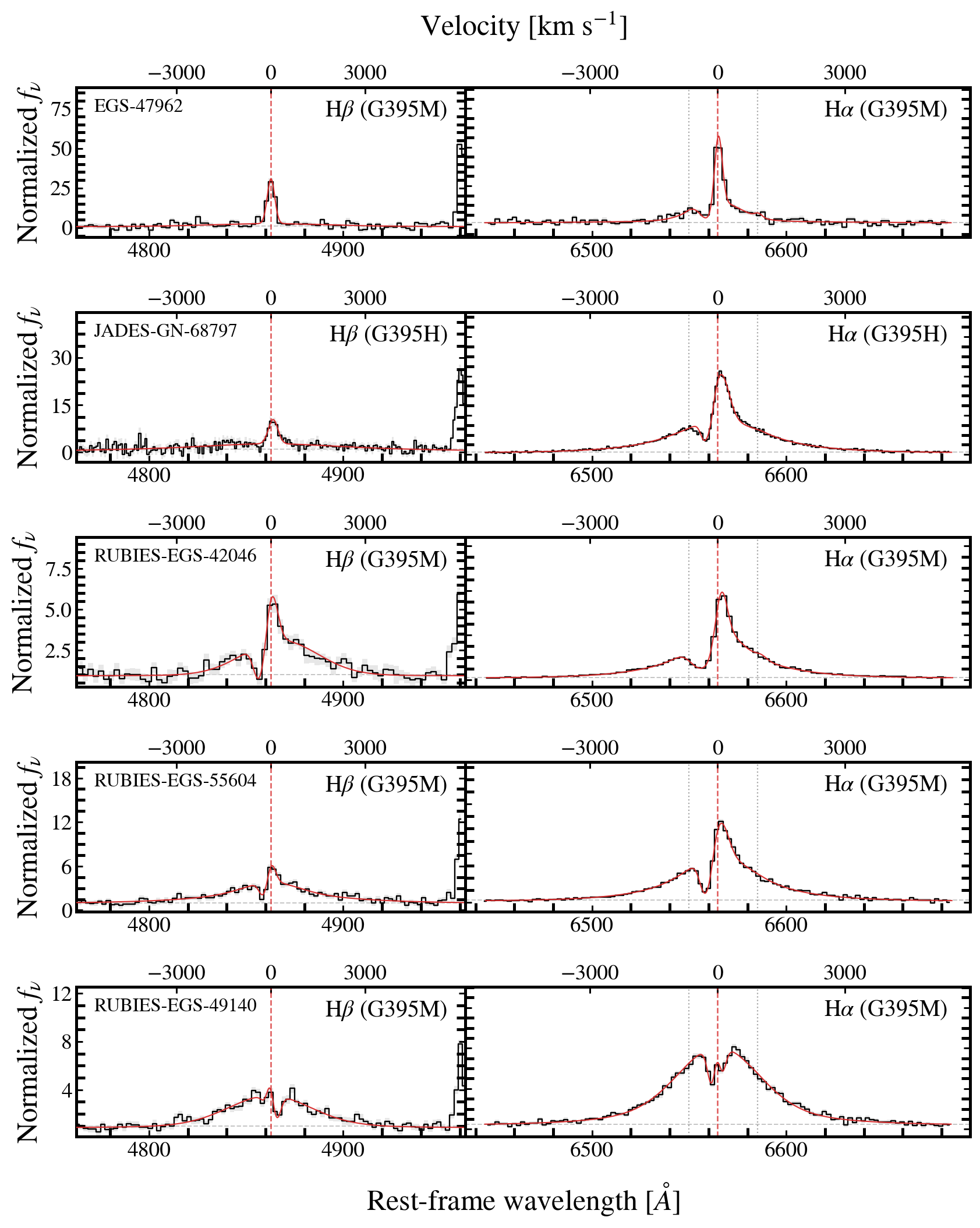}
    \caption{JWST/NIRSpec medium- or high-resolution grating spectra of the 14 LRDs with detected H$\alpha$ absorption. For each object the H$\beta$ panel (left) is shown together with the H$\alpha$ panel (right) on a common rest-wavelength scale. Black lines show the observed spectra, gray shaded regions show the 1$\sigma$ uncertainties, red curves show the best-fit line-profile models, red dashed vertical lines mark H$\alpha$ and H$\beta$ at the systemic redshift measured with [O\,\textsc{iii}]$\lambda5007$, and gray dotted vertical lines mark [N\,\textsc{ii}]$\lambda\lambda6548,6583$.}
    \label{fig:sample_spectra}
\end{figure*}

\begin{figure*}
    \centering
    \includegraphics[width=1\linewidth]{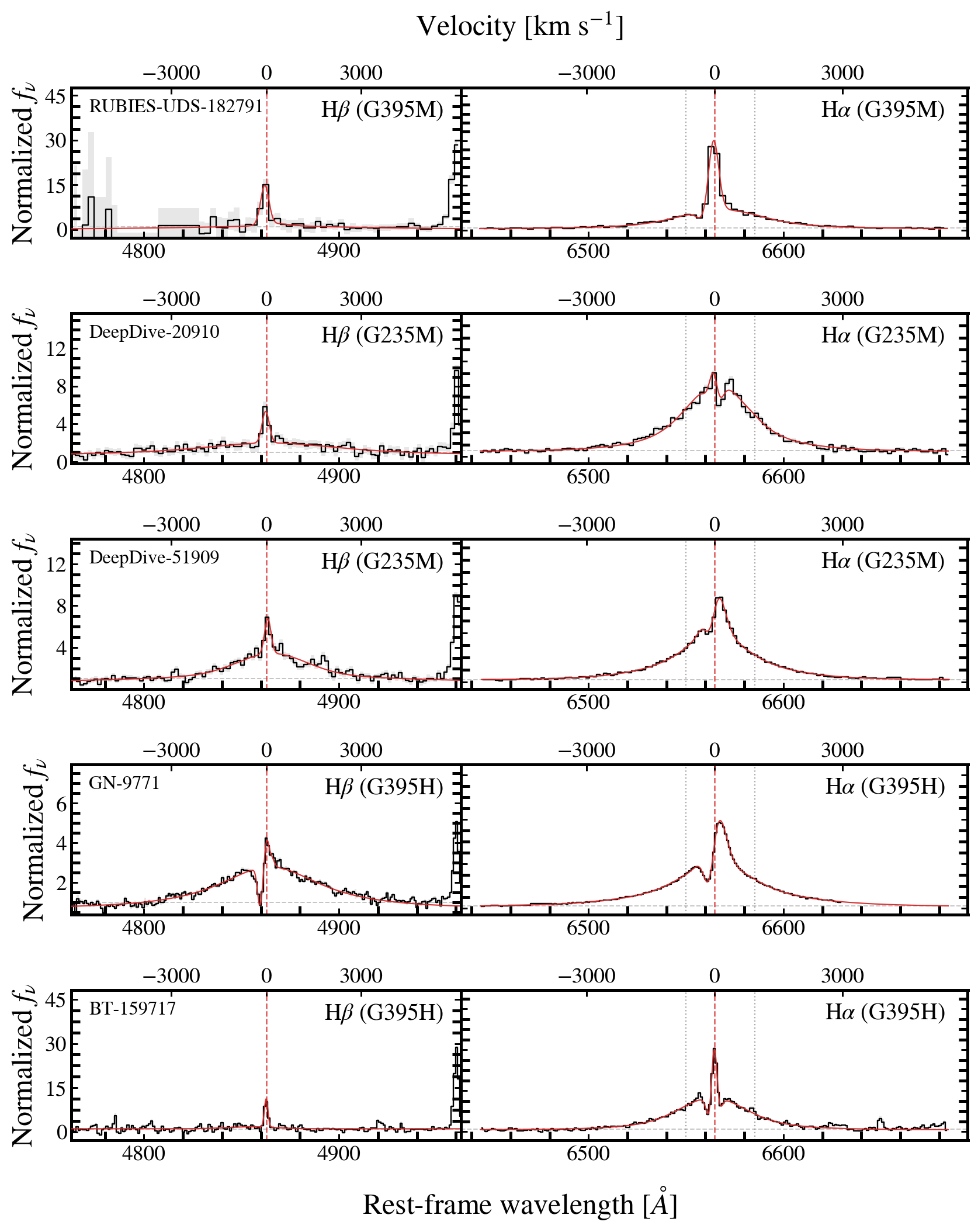}
    \caption{Same as Figure~\ref{fig:sample_spectra} for the next five LRDs with detected H$\alpha$ absorption.}
    \label{fig:sample_spectra_cont1}
\end{figure*}

\begin{figure*}
    \centering
    \includegraphics[width=1\linewidth]{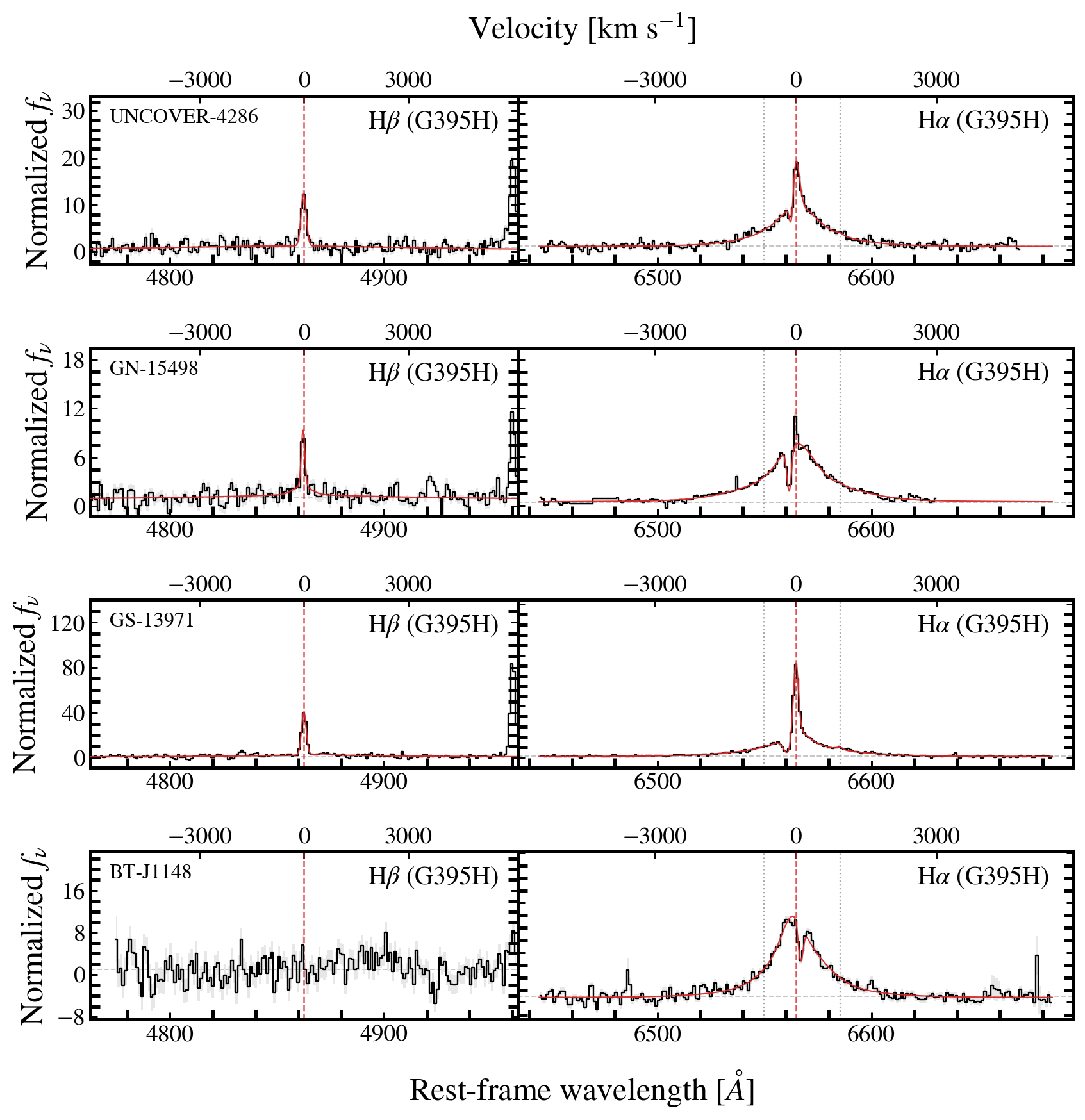}
    \caption{Same as Figure~\ref{fig:sample_spectra} for the remaining four LRDs with detected H$\alpha$ absorption.}
    \label{fig:sample_spectra_cont2}
\end{figure*}

\section{Sample and Data}\label{sec:data}
Our analysis builds on the compact broad-H$\alpha$ sample construction of \citet{Yanagisawa+2026}, but with two updates for the present absorption-search sample: we do not require PRISM coverage, and we classify LRDs using the photometry and compactness measurements described below. We refer the reader to \citet{Yanagisawa+2026} for the full description of the data reduction, continuum measurements, broad-line luminosity measurements, and the other aspects of the parent-sample selection.

The parent sample combines publicly available JWST/NIRSpec spectra from the DAWN JWST Archive (DJA; \citealt{Heintz+2024, deGraaff+2025a, Valentino+2025}) version~4.4\footnote[1]{\url{https://zenodo.org/records/15472354}} with complementary NIRSpec/IFU observations from GO 5664 (PI: Matthee) and GO 5015 (BlackTHUNDER; PI: {\"U}bler and Maiolino). Briefly, the DJA selection starts from sources with reliable spectroscopic redshifts (grade 3) and medium- or high-resolution grating coverage of both H$\alpha$ and [O\,\textsc{iii}]$\lambda5007$ with peak signal-to-noise ratio larger than 5 for line measurements; PRISM spectra are not required for the sample selection. Broad-H$\alpha$ sources are then selected using the broad-line criteria described in \citet{Yanagisawa+2026}. Sources position-matched within $1\arcsec$ are counted once: for multiple MSA observations, we adopt the spectrum with the highest broad-H$\alpha$ signal-to-noise ratio, while for objects present in both the DJA and IFU data sets, we use the higher-S/N IFU spectrum.

We select LRDs from the broad-H$\alpha$ parent sample using imaging-based photometry and compactness measurements. For the DJA objects, we perform $0.^{\prime\prime}2$ aperture photometry on the DJA imaging products, whose point spread functions are matched to that of the F444W image, and apply the color selection shown in Figure~\ref{fig:color}. For a source at redshift $z$, the adopted cuts are
\begin{equation}\label{eq:color}
\begin{aligned}
z<4: \quad & m_{\rm F150W}-m_{\rm F277W}>1.0, \\
           & m_{\rm F090W}-m_{\rm F115W}<1.0, \\
4<z<6: \quad & m_{\rm F200W}-m_{\rm F356W}>1.0, \\
           & m_{\rm F115W}-m_{\rm F150W}<1.0, \\
z>6: \quad & m_{\rm F277W}-m_{\rm F444W}>1.0, \\
           & m_{\rm F150W}-m_{\rm F200W}<1.0,
\end{aligned}
\end{equation}
where $m_{\rm filter}$ denotes the AB magnitude in each NIRCam filter. These cuts identify sources with blue rest-UV colors and red rest-optical colors. We note that these selections are arbitrary to some extent, as several previous studies suggest spectral diversity in LRDs may reflect the various contributions from the central engine and its host galaxy \citep[e.g.,][]{Perez-Gonzalez+2026, Sun+2026}. We adopt the F444W compactness criterion of \citet{Akins+2025}, $C_{444}\equiv F_{0\farcs2}/F_{0\farcs5}>0.5$, where $F_{0\farcs2}$ and $F_{0\farcs5}$ are measured in circular apertures with diameters of $0\farcs2$ and $0\farcs5$, respectively. Objects failing either the color or compactness criterion are excluded. 
%We further require medium- or high-resolution grating coverage of the H$\alpha$ region and the [O\,\textsc{iii}]$\lambda5007$ systemic-redshift tracer. 
However, JADES-GN-68797 is the sole exception to the compactness cut. It exhibits unambiguous H$\alpha$ absorption but has $C_{444}=0.398$, below the adopted threshold, because its compact core is accompanied by an extended component. \citet{Setton+2025b} likewise identify diffuse emission while noting a clearly compact core, and its V-shaped SED and spectroscopic properties are characteristic of LRDs. We therefore retain this source in the LRD sample.
For BT-J1148, the available imaging does not provide all of the bands needed for Equation~\eqref{eq:color}, but a PRISM spectrum is available. We therefore compute synthetic photometry from the PRISM spectrum and apply the same color cuts. Although PRISM coverage is not generally required, it is used here only to replace the missing imaging photometry. Because F444W imaging is also unavailable, we assess compactness in the available F356W image using the same aperture diameters and find $C_{356}=0.65$, above the adopted threshold. The resulting absorption-search sample comprises 40 LRDs spanning $z_{\rm spec}\simeq2.5$--7.2.

\section{Line Profile Fitting}\label{sec:line_profile_fitting}
We model the medium- and high-resolution grating spectra following the emission-line fitting framework of \citet{Yanagisawa+2026}, but with an updated treatment of Balmer absorption. In brief, we fit the H$\alpha$+[N\,\textsc{ii}] complex with a linear continuum, a narrow H$\alpha$ Gaussian component with $\mathrm{FWHM}<500\,\mathrm{km\,s^{-1}}$, a broad H$\alpha$ component modeled as an intrinsic Gaussian, a fraction $f_{\rm scatt}$ of which is redistributed into exponential wings by electron scattering with a velocity scale $w_{\rm scatt}$ \citep[following][]{Scholtz+2026}, [N\,\textsc{ii}]$\lambda\lambda6548,6583$ Gaussian components with their flux ratio fixed to the theoretical value of 2.94, and, when warranted by the data, a partial-covering H$\alpha$ absorption component. Unlike \citet{Yanagisawa+2026}, where the absorption was represented by an additive Gaussian component, we model the absorption as a multiplicative transmission applied to the continuum plus broad H$\alpha$ component:
\begin{equation}
\begin{split}
T(\lambda) &= 1-C_f+C_f\exp[-\tau(\lambda)],\\
\tau(\lambda) &= \tau_0\exp\left[-\frac{(v-v_{\rm abs})^2}{2\sigma_{\rm abs}^2}\right].
\end{split}
\end{equation}
The absorption centroid $v_\mathrm{abs}$, width $\sigma_\mathrm{abs}$, central optical depth $\tau_0$, and covering factor $C_f$ are free parameters. The absorbed background and emission-line components are convolved with the NIRSpec line-spread functions from \citet{Isobe+2023}. 
When H$\beta$ is available, we fit it with the same partial-covering prescription, except that no [N\,\textsc{ii}] doublet is included. The H$\beta$ absorption is fitted independently of H$\alpha$. 
From the best-fit transmission profile, we measure the rest-frame equivalent width of the absorption as $\mathrm{EW_{abs}}=\int[1-T(\lambda)]\,d\lambda/(1+z)$, and derive the column density of hydrogen in the $n=2$ level as $N_{n=2}=\frac{m_{\rm e}c}{\pi e^{2}f\lambda_{0}}\int\tau(v)\,dv$, where $f=0.6407$ is the H$\alpha$ oscillator strength and $\lambda_{0}$ is the rest-frame wavelength of H$\alpha$. 
%Because this paper focuses on Balmer absorption, the key quantity is whether the partial-covering absorption component is statistically required. % and, if so, its velocity offset. 
We fit each object with and without the absorption component and compute $\Delta\mathrm{BIC}\equiv\mathrm{BIC}_{\rm no\,abs}-\mathrm{BIC}_{\rm abs}$. We adopt the absorption-included solution when $\Delta\mathrm{BIC}>10$ and otherwise retain the no-absorption model. We then visually inspect the statistically selected candidates and reject cases in which the absorption model is favored only because it accommodates an asymmetric emission-line profile but no local flux minimum (i.e., no trough-like feature) is present. The systemic redshift is measured from [O\,\textsc{iii}]$\lambda5007$, and the absorption velocity offsets, $\dvabs(\ha)$ and $\dvabs(\hb)$, are defined relative to this systemic frame. The line-profile fits for the detected absorbers are shown in Figures~\ref{fig:sample_spectra}--\ref{fig:sample_spectra_cont2}. The measured properties of broad emission and absorption are listed in Tables \ref{tab:balmer_emission_properties} and \ref{tab:balmer_abs_properties}, respectively.
For completeness, Tables~\ref{tab:balmer_emission_properties} and \ref{tab:balmer_abs_properties} also compile H$\alpha$ emission and absorption properties of Balmer absorbers reported in previous studies. This compilation is not homogeneous: the contributing studies differ in sample selection, spectral resolution, line-profile decomposition, and systemic-redshift tracer. The tables therefore retain every reported study, observing epoch, and resolved absorption component rather than attempting to homogenize the measurements. Nevertheless, because the number of known Balmer-absorption LRDs remains small, it is worthwhile to assemble and compare as many of the available measurements as possible, even at the expense of full homogeneity. In the population-level analyses, repeated epochs of the same physical absorption component are represented by the median of the reported measurements, whereas kinematically distinct absorption components in a single object are retained as separate absorbers; when counting host objects rather than absorbers, such a system is counted only once. We adopt these conventions throughout. JADES-GN-28074 and JADES-GN-38147, although reported as LRD Balmer absorbers in previous studies \citep{Juodzbalis+2026}, fail the red rest-optical color cuts adopted here. We therefore exclude them from our uniformly selected sample but retain their published measurements in the literature compilation and comparison plots.

\section{Results}\label{sec:results}

\subsection{Absorber Fraction}
\begin{figure}
    \centering
    \includegraphics[width=1\linewidth]{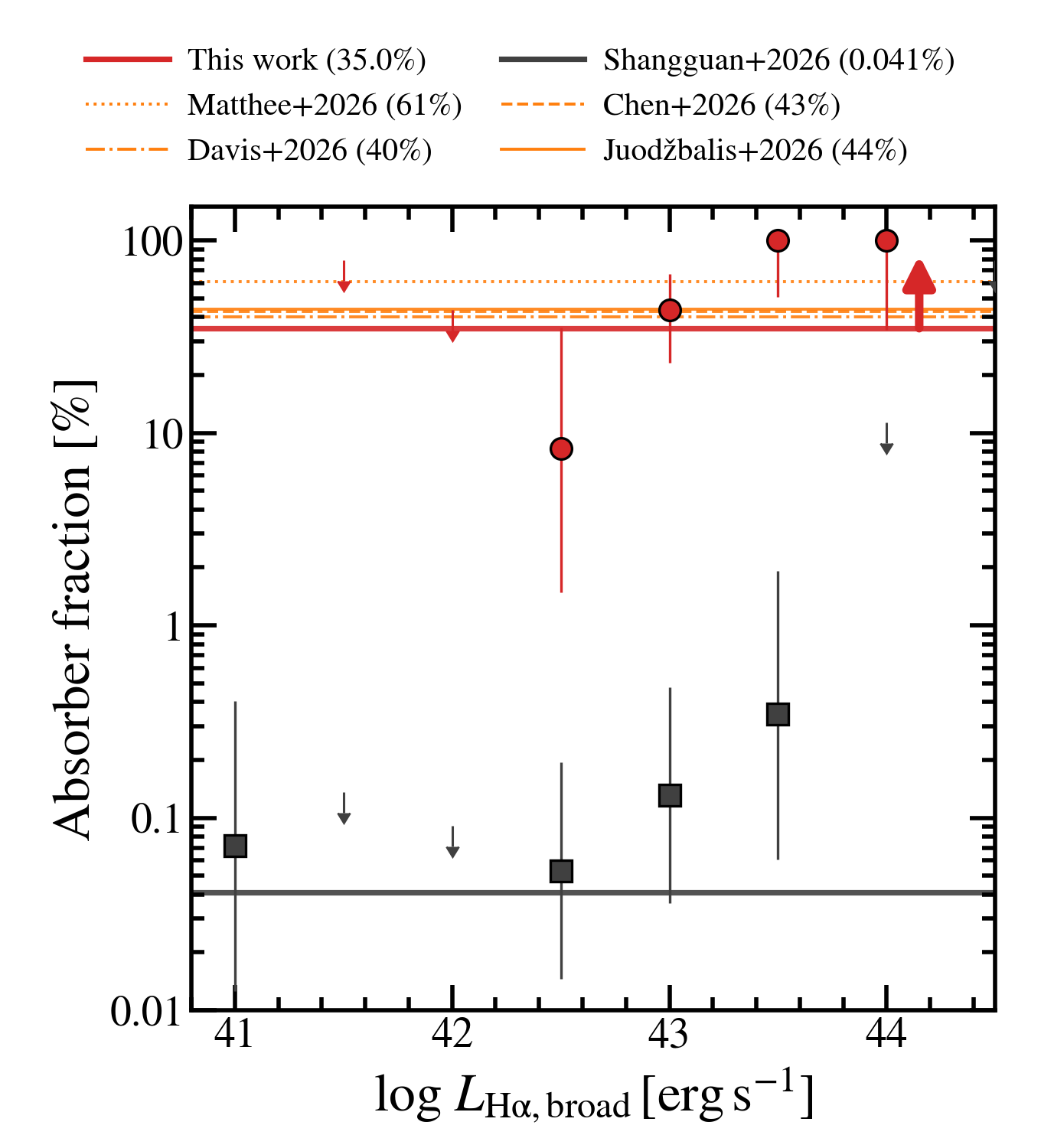}
    \caption{Balmer-absorption fraction for LRDs and low-$z$ type 1 AGNs.  Red circles show the LRD sample in this work, and black squares show the SDSS type 1 AGN comparison sample constructed from \citet{Shangguan+2026}. Error bars show 95\% Wilson confidence intervals; downward arrows denote one-sided upper limits for bins with zero detected absorbers. The solid red and black lines show the observed overall fraction for LRDs and SDSS type 1 AGNs. Orange dotted, dashed, and dash-dotted lines mark the observed fractions reported by \citet{Matthee+2026}, \citet{Chen+2026}, and \citet{Davis+2026}, respectively; the orange solid line marks the completeness-corrected fraction inferred by \citet{Juodzbalis+2026}.}
    \label{fig:abs_frac}
\end{figure}

Figure \ref{fig:abs_frac} shows the absorber fraction (the number of objects with Balmer absorption divided by the total number of objects) as a function of broad H$\alpha$ luminosity $\lbha$. We detect Balmer absorption in $14/40=35\%$ of the LRDs, compared with $6/14583\simeq0.04\%$ of the SDSS type 1 AGNs \citep[][]{Shangguan+2026}, a difference of a factor of approximately 850. Seven of the 14 LRD absorbers are detected with medium-resolution gratings ($R\sim1000$), whose spectral resolution is lower than that of SDSS ($R\sim2000$); the difference in absorber fraction therefore cannot be explained solely by spectral resolution.

Our measured fraction should be regarded as an observed lower limit rather than a completeness-corrected intrinsic incidence. Using fully synthetic broad-H$\alpha$ spectra spanning a wide range of absorber properties, \citet{Juodzbalis+2026} showed that JWST/NIRSpec observations---particularly at $R\sim1000$---can miss a substantial fraction of Balmer absorption.
%, especially when an absorber lies near the systemic velocity and is filled in by narrow H$\alpha$ emission. 
They inferred a completeness-corrected H$\alpha$-absorption incidence of $44^{+21}_{-6}\%$ for LRDs. Because their parent-sample definition, signal-to-noise selection, and distribution of spectral sensitivities differ from ours, we do not apply this correction directly to our sample. Nevertheless, incompleteness on the LRD side would increase rather than reduce the contrast with the observed SDSS type~1 AGN fraction. 

\subsection{Velocity Offset Distributions}
\begin{figure*}
    \centering
    \includegraphics[width=1\linewidth]{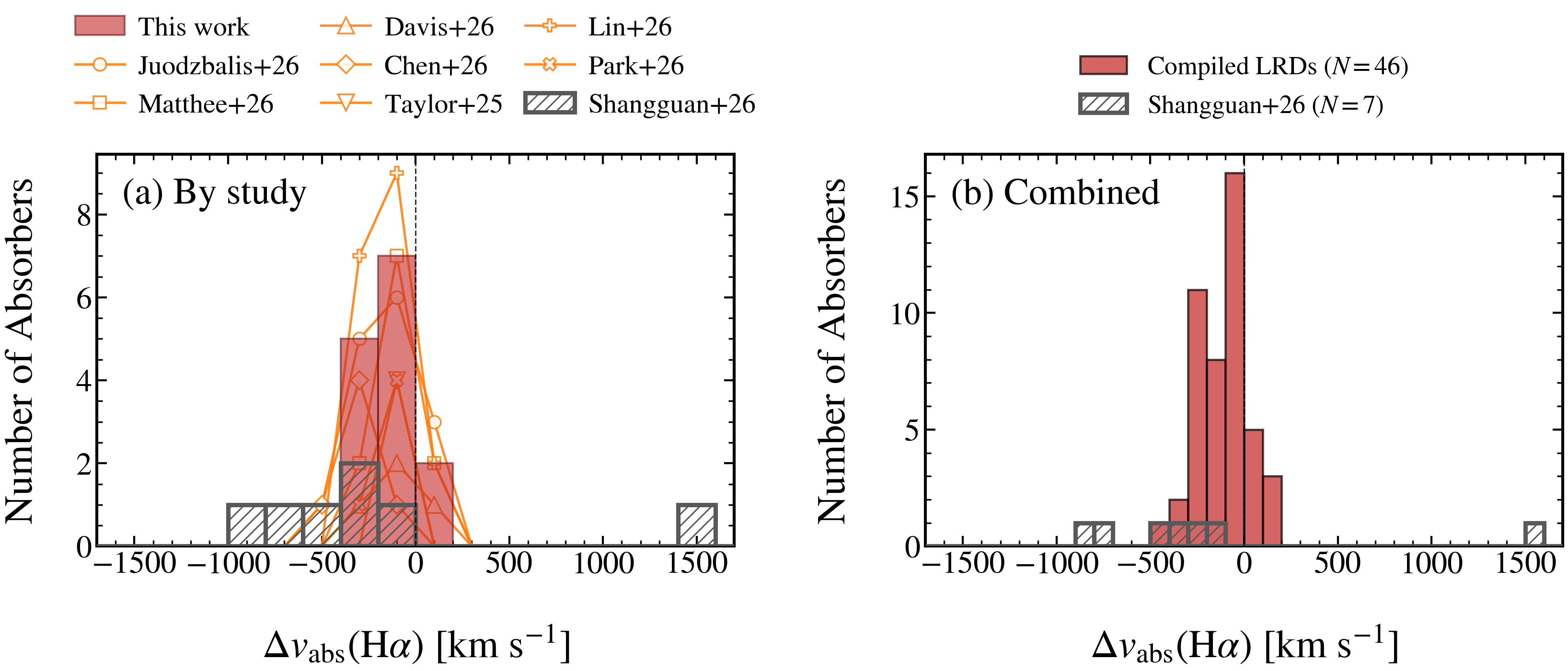}
    \caption{Distribution of the H$\alpha$ absorption velocity offset. (a) Measurements separated by study. The filled red histogram shows this work; orange frequency polygons with distinct markers show the bin counts from \citet{Taylor+2025}, \citet{Matthee+2026}, \citet{Davis+2026}, \citet{Chen+2026}, \citet{Juodzbalis+2026}, \citet{XLin+2026}, and \citet{Park+2026}; and the hatched gray histogram shows the low-$z$ SDSS type~1 AGNs of \citet{Shangguan+2026}. Panel (a) retains measurements reported independently by different studies even when they refer to the same object. (b) Compilation of the absorbers in this work and literature absorbers not included in our sample; the selection of one measurement for multiply reported literature objects follows the criteria described in the text. The combined LRD compilation contains 46 absorption components in 45 unique objects, while the SDSS comparison contains seven components in six unique objects.}
    \label{fig:abs_hist}
\end{figure*}

Figure~\ref{fig:abs_hist} shows the H$\alpha$ absorption velocity offsets, $\dvabs(\ha)$, measured in this work and literature. In our sample, $12/14=86\%$ of the H$\alpha$ absorbers are blueshifted, with a median $\dvabs(\ha)=-117\,\mathrm{km\,s^{-1}}$ and a range from $-375$ to $+100\,\mathrm{km\,s^{-1}}$. The same qualitative concentration toward modest blueshifts is seen in independent high-redshift samples \citep{Taylor+2025, Matthee+2026, Davis+2026, Chen+2026, Juodzbalis+2026} and in low-redshift LRD samples \citep{XLin+2026, Park+2026}. To construct the combined sample, we supplement the absorbers in this work with literature objects that are not included in our sample. When such an object has been reported by multiple studies, we select one measurement with higher-resolution spectroscopy with a systemic redshift anchored by [O\,\textsc{iii}]% or another narrow-line tracer
; where these criteria do not distinguish among measurements, we adopt the result from the more recent uniform analysis. 
After matching sources by coordinates and applying this selection, the combined sample contains 46 absorption components in 45 unique objects; $38/46=83\%$ are blueshifted, with a median $\dvabs(\ha)=-95\,\mathrm{km\,s^{-1}}$ and a range from $-425$ to $+137\,\mathrm{km\,s^{-1}}$. Thus, the prevalence of low-velocity blueshifted absorption is not specific to our fitting analysis. The SDSS type~1 AGNs of \citet{Shangguan+2026} are likewise predominantly blueshifted ($6/7$ components), but span a much broader interval, from $-845$ to $+1587\,\mathrm{km\,s^{-1}}$.

\begin{figure}
    \centering
    \includegraphics[width=1\linewidth]{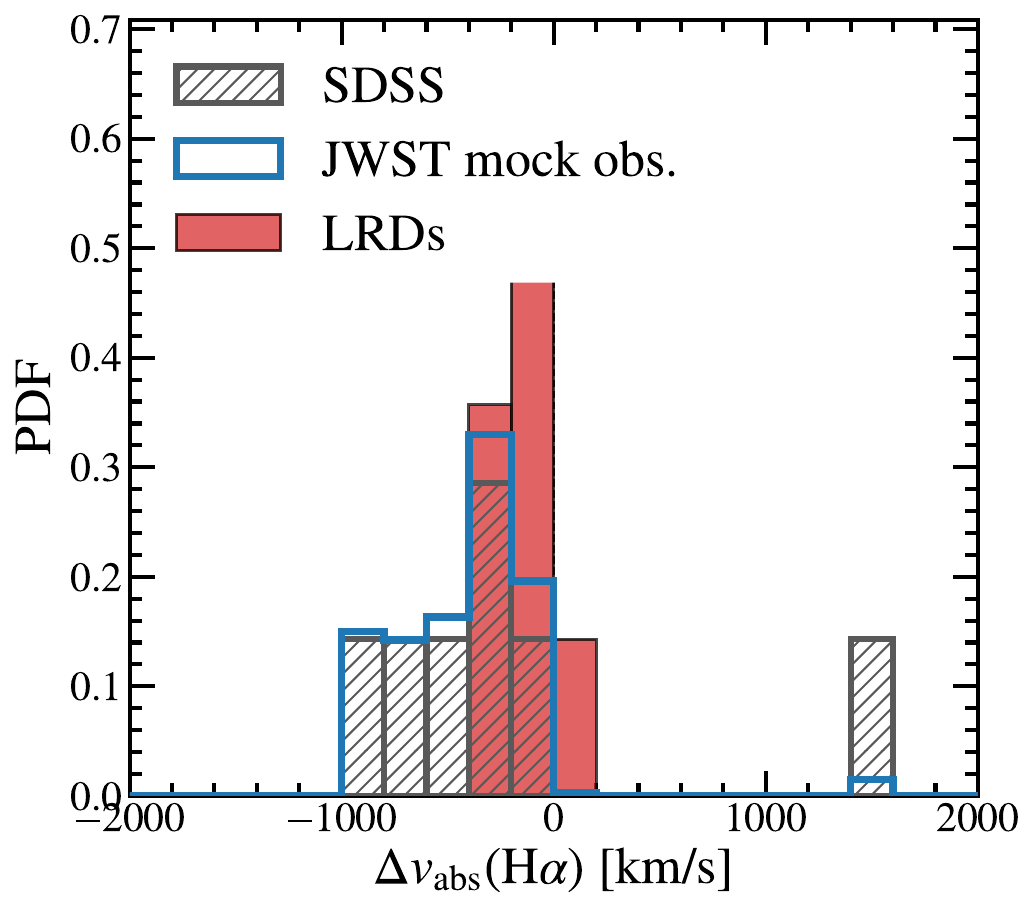}
    \caption{Forward modeling of the H$\alpha$ absorption velocity distribution. The histogram shows the injected empirical SDSS absorption-profile (hatched gray histogram), the mock observed distribution (blue outline), and the observed LRD absorbers (red histogram).}
    \label{fig:velocity_forward_model}
\end{figure}

To determine whether this velocity difference could arise from the heterogeneous sensitivity of the LRD spectra, we perform a semi-synthetic injection--recovery experiment using the empirical SDSS Balmer-absorption profiles from \citet{Shangguan+2026}. For each LRD spectrum, we reconstruct the adopted no-absorption H$\alpha$+[N\,\textsc{ii}] model obtained in this work. We then inject an SDSS absorption transmission profile into the continuum plus broad H$\alpha$ component before line-spread-function convolution, while leaving narrow H$\alpha$ and [N\,\textsc{ii}] unabsorbed, and add Gaussian noise drawn from the observed error spectrum. 
%The SDSS Balmer line absorber in \cite{Shangguan+2026} contains seven absorption components and retains their joint distribution of $\dvabs(\ha)$, equivalent width, line width, and optical depth.
We generate 50 independent noise realizations for every LRD--profile pairing and refit every mock spectrum with the same no-absorption and partial-covering absorption models and the same $\Delta\mathrm{BIC}>10$ criterion used for the data, while allowing the absorption center to vary over $-3000<\dvabs(\ha)<3000\ \mathrm{km\,s^{-1}}$. We use these recovery probabilities to conduct 1,000,000 mock surveys, assigning one empirical SDSS profile and one noise realization to each of the 40 LRD spectra. 
%The intrinsic incidence of Balmer absorption does not enter the velocity-shape test: 
We condition the comparison on mock surveys yielding exactly 14 recovered absorbers, matching the number of recovered LRD absorbers with the observed one. 
A total of 7,114 mock surveys satisfy this condition. Figure~\ref{fig:velocity_forward_model} compares the SDSS profile distribution, the observed LRD distribution, and the mock observed distribution. The selection function preferentially removes some SDSS profiles, especially those at $|\dvabs|>1000\,\mathrm{km\,s^{-1}}$, but does not compress their velocities sufficiently to reproduce the LRD measurements: the mock detected samples have a median $\dvabs(\ha)=-393\ \mathrm{km\,s^{-1}}$ 
%and a median velocity dispersion of $247\ \mathrm{km\,s^{-1}}$, 
compared with $-117\ \mathrm{km\,s^{-1}}$ 
%and $126\ \mathrm{km\,s^{-1}}$ 
for the observed LRDs. Thus, within the assumptions of the empirical SDSS profile library and our semi-synthetic spectra, observational detectability alone does not explain the narrow LRD velocity distribution.
%; the evidence is strongest in the high-resolution subsample and remains significant in the medium-resolution subsample.

\begin{figure}
    \centering
    \includegraphics[width=1\linewidth]{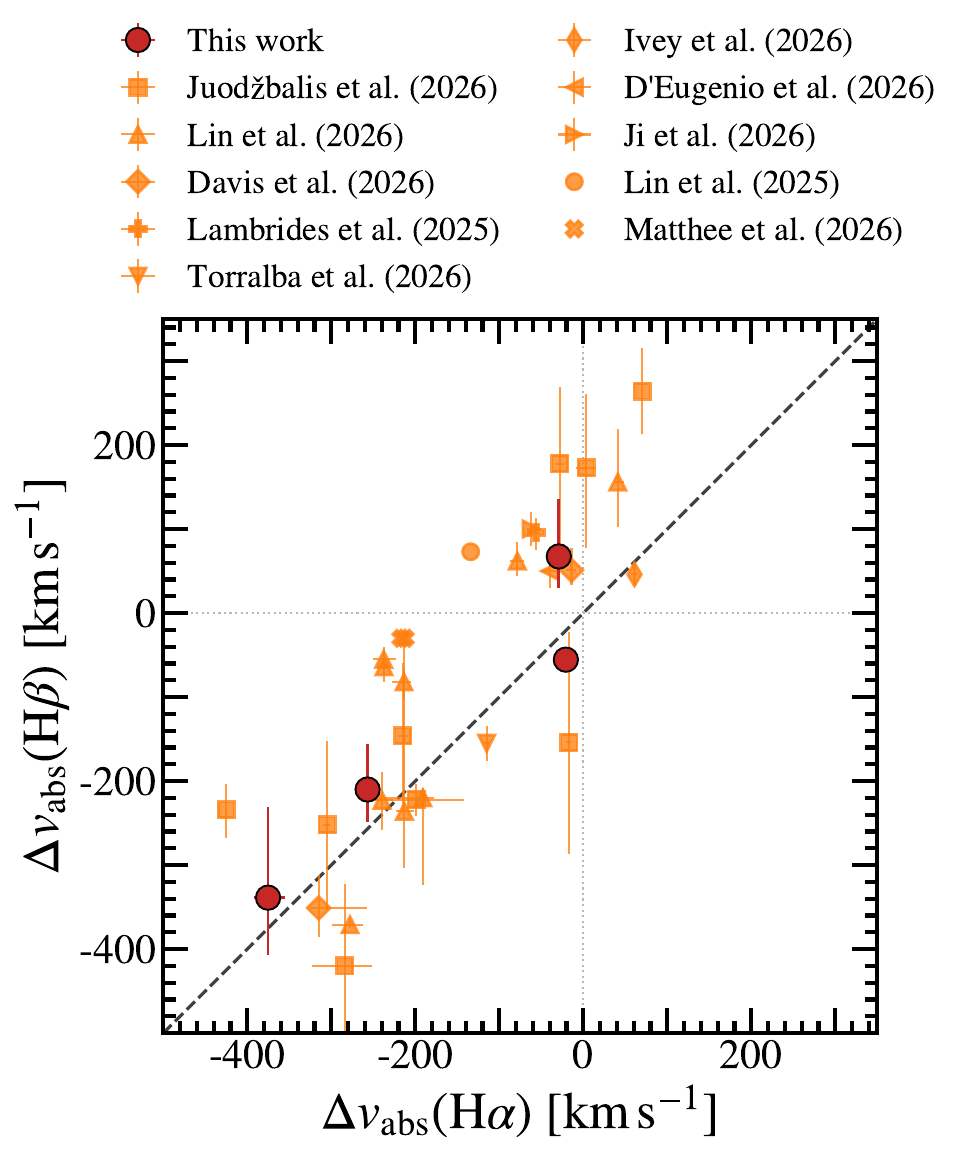}
    \caption{H$\beta$ versus H$\alpha$ absorption velocity offsets. Red circles show this work; orange symbols show literature measurements from \citet{Juodzbalis+2026}, \citet{XLin+2026}, \citet{Davis+2026}, \citet{LambridesAbs+2025}, \citet{Torralba+2026}, \citet{Ivey+2026}, \citet{DEugenioQSO1+2026}, \citet{JiLRD+2026}, \citet{XLin+2025}, and \citet{Matthee+2026}. All 32 measurements from 11 studies are plotted, including repeated measurements of the same physical object by different studies; these measurements represent 19 physical objects. The dashed line marks equal H$\alpha$ and H$\beta$ offsets, and the dotted lines mark zero velocity.}
    \label{fig:halpha_hbeta_dv}
\end{figure}

We detect H$\beta$ absorption in four of the 14 H$\alpha$ absorbers. Figure~\ref{fig:halpha_hbeta_dv} compares their H$\alpha$ and H$\beta$ velocity offsets with a compilation of published measurements for LRDs. The plot contains all 32 measurements from 11 studies, representing 19 physical objects; repeated measurements of the same object by different studies are deliberately retained. In three of the four objects in this work, the two centroids agree to within $\simeq50\,\mathrm{km\,s^{-1}}$; RUBIES-EGS-49140 instead differs by $\simeq96\,\mathrm{km\,s^{-1}}$ and has opposite-sign centroids. The literature measurements show substantially greater object-to-object diversity. The recovered centroids can differ because the two transitions have different optical depths and reflect kinematic information at different cloud depth. The absorption profiles are superposed on different combinations of narrow and broad emission between $\ha$ and $\hb$, making the inferred absorption center sensitive to emission infill and line-profile decomposition \citep[e.g.,][]{Matthee+2026, Davis+2026}. An intriguing asymmetry emerges among the eight measurements of four physical objects with opposite-sign Balmer centroids: all exhibit blueshifted H$\alpha$ and redshifted H$\beta$, whereas the converse configuration is absent. Because both transitions arise from the $n=2$ level, their different optical depths alone do not select a velocity sign; however, the weaker H$\beta$ transition preferentially retains components with larger $n=2$ column density or, if the effective covering is transition dependent, a larger covering of the H$\beta$ background. If the two transitions consequently weight different radii in a stratified, partially covered medium, the observed orientation is consistent with a dense inner component dominated by inflow and a lower-column outer component dominated by outflow, broadly similar to interpretations proposed for individual LRDs \citep{DEugenioIrony+2025, Davis+2026, JiLRD+2026}. %Alternatively, a common transition-dependent bias caused by partial covering, differential emission infill, or decomposition of the distinct underlying line profiles could preferentially shift the recovered H$\beta$ centroid redward relative to H$\alpha$ \citep{Juodzbalis+2026, Davis+2026}. 
Given only four independent objects and the heterogeneous measurements, we regard this one-sided pattern as suggestive rather than decisive.

\subsection{Equivalent Widths of Absorption and Emission}
\begin{figure*}
    \centering
    \includegraphics[width=1\linewidth]{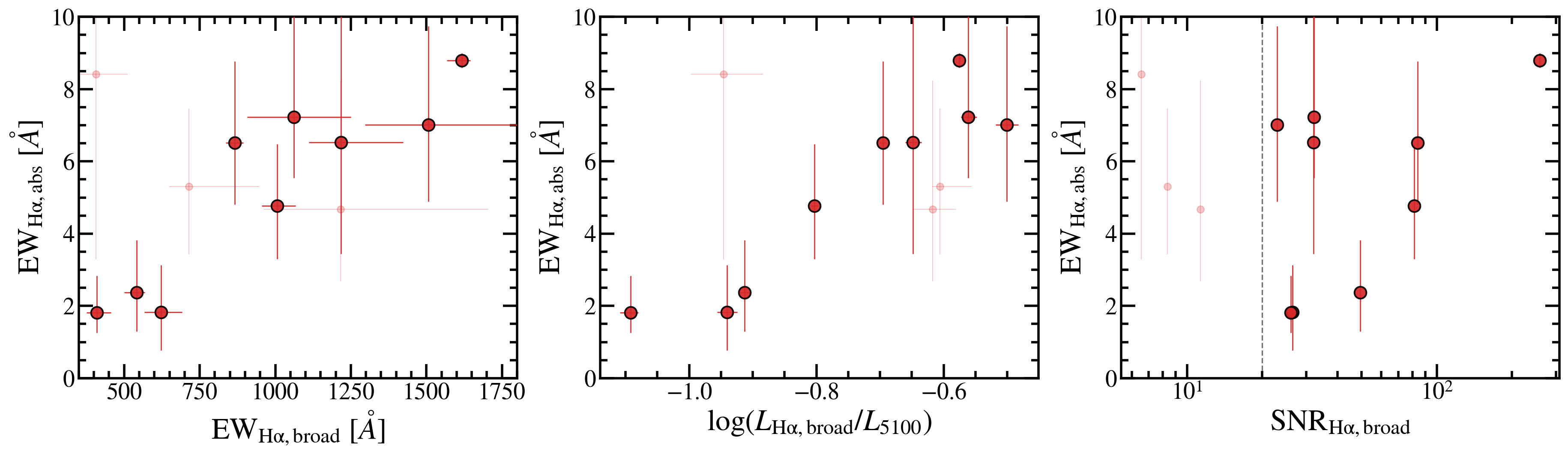}
    \caption{Rest-frame H$\alpha$ absorption equivalent width, ${\rm EW}_{\mathrm{H}\alpha,\mathrm{abs}}$, as a function of the rest-frame broad-H$\alpha$ emission equivalent width (left), the broad-H$\alpha$-to-5100~\AA\ continuum luminosity ratio, $\log(L_{\mathrm{H}\alpha,\mathrm{broad}}/L_{5100})$ (middle), and the signal-to-noise ratio of the broad H$\alpha$ component (right). Only the 12 absorbers with an available $L_{5100}$ measurement are shown. Red circles denote the nine LRDs with $\mathrm{S/N}_{\mathrm{H}\alpha,\mathrm{broad}}>20$ used in the correlation tests. Small red circles denote the three objects with $\mathrm{S/N}_{\mathrm{H}\alpha,\mathrm{broad}}\leq20$, whose equivalent-width measurements are treated as incomplete and excluded from the correlation tests. The vertical dashed line in the right panel marks $\mathrm{S/N}_{\mathrm{H}\alpha,\mathrm{broad}}=20$.}
    \label{fig:ew}
\end{figure*}

The left panel of Figure~\ref{fig:ew} compares the H$\alpha$ absorption and broad-emission equivalent widths. To mitigate potential incompleteness in weak absorption at low broad-H$\alpha$ signal-to-noise ratio, we restrict the correlation analysis to the nine objects with $\mathrm{S/N}_{\ha,\mathrm{broad}}>20$ among the 12 absorbers with an available $L_{5100}$ measurement; the three lower-S/N objects are shown but excluded from the test. The high-S/N subset shows a strong correlation between $\mathrm{EW}_{\ha,\mathrm{abs}}$ and $\mathrm{EW}_{\ha,\mathrm{broad}}$ (Spearman $\rho=0.917$, $p=5.07\times10^{-4}$). Within the high-S/N subset, the absorption equivalent width shows no significant correlation with the broad-H$\alpha$ signal-to-noise ratio (Spearman $\rho=0.350$, $p=0.356$; right panel), indicating that broad-H$\alpha$ signal-to-noise alone is unlikely to explain the correlations in the left and middle panels, although the current sample remains small. 
To examine the physical origin of this trend, we replace the broad-H$\alpha$ equivalent width with $L_{\mathrm{H}\alpha,\mathrm{broad}}/L_{5100}$ (the middle panel of Figure~\ref{fig:ew}) and again find a strong positive correlation (Spearman $\rho=0.933$, $p=2.36\times10^{-4}$; middle panel).
Note that these two quantities are not independent because both express the broad-H$\alpha$ line strength relative to the optical continuum; the middle panel should therefore be regarded as a recasting. Nevertheless, $L_{\mathrm{H}\alpha,\mathrm{broad}}/L_{5100}$ serves as a proxy for the offset of an LRD from the broad-H$\alpha$--continuum luminosity relation of low-$z$ type~1 AGNs: \citet{Yanagisawa+2026} showed that LRDs have enhanced broad H$\alpha$ emission at fixed continuum luminosity and argued that the enhancement becomes larger as the covering factor and column density of the BLR gas increase. In this picture, the correlations in the left and middle panels are qualitatively consistent with a common increase in the amount and geometrical covering of dense gas in and around the BLR, which can both enhance broad-H$\alpha$ emission and produce stronger H$\alpha$ absorption along the observed line of sight. %Although the correlations are statistically significant within the adopted high-S/N subset, this physical interpretation remains tentative because the sample is small and the two quantities shown on the horizontal axes are not independent. %We emphasize, however, that the absorption equivalent width is not a direct measure of the total gas mass, because it also depends on the $n=2$ population, optical depth, covering factor, and velocity width.
%A possible observational concern is that stronger absorption may be preferentially recovered when the broad H$\alpha$ component is detected at higher significance. However, 

\section{Discussion}\label{sec:discussion}

\subsection{Incidence and Geometry of the Dense Gas}
The approximately 850-fold contrast in the observed Balmer-absorption incidence indicates that sight lines to the compact continuum and broad-line emitting regions intersect dense, excited hydrogen much more frequently in LRDs than in low-redshift type~1 AGNs. This population incidence should not, however, be identified directly with a global solid-angle covering factor: it also depends on the absorber geometry and duty cycle, as well as on detection completeness. Likewise, it is distinct from the fitted partial-covering parameter $C_f$, which measures the fraction of the continuum plus broad-H$\alpha$ background occulted along an individual detected sight line. Nevertheless, the high incidence, together with $C_f\gtrsim0.7$ for the detected absorbers (Table~\ref{tab:balmer_abs_properties}), is qualitatively consistent with scenarios in which the nuclei of LRDs are surrounded by dense, high-column-density gas with a larger effective covering probability than in ordinary type~1 AGNs \citep[e.g.,][]{Inayoshi_Maiolino_2025, Naidu+2025, deGraaff+2025b, Ji+2025, Rusakov+2026, Sneppen+2026a, Pacucci+2026, Yanagisawa+2026}.

For example, \cite{Schulze+2018} have identified Balmer absorption lines in two quasars and derived moderately high covering factors of $C_f=0.2$--0.4 in several scenarios. Such high column density gas may produce strong nebular continuum mimicking the modified-blackbody in the rest-frame optical \citep{Rusakov+2026, Sneppen+2026a, Sneppen+2026b, Pacucci+2026, Ando+2026,Yanagisawa+2026}. The absence of similarly red optical continua in low-redshift type~1 AGNs with Balmer absorption may therefore reflect a smaller effective solid-angle coverage and/or a smaller local covering of the continuum-emitting region, rather than the absence of dense gas altogether.

\subsection{Absorber Kinematics Relative to the BLR Dynamical Scale}
The physical origin of the velocity difference between LRDs and low-redshift type~1 AGNs is unclear. A useful first question is whether the lower absolute velocities in LRDs track a smaller characteristic dynamical velocity in their nuclear regions. 
%, as indicated by their generally narrower broad lines. 
This comparison is motivated by AGN wind models and UV/X-ray absorber studies in which outflow velocities scale with the escape or rotational velocity at the launch radius \citep[e.g.,][]{Proga+2000, Laor_Brandt_2002, Fukumura+2010}. %Here we use the escape velocity only as a normalization of the BLR dynamical velocity scale; once radiation pressure is included, $|\dvabs(\ha)|/v_{\rm esc}$ alone is not a criterion for whether an absorber is gravitationally bound (Section \ref{sec:radiation}). 

\begin{figure}
    \centering
    \includegraphics[width=1\linewidth]{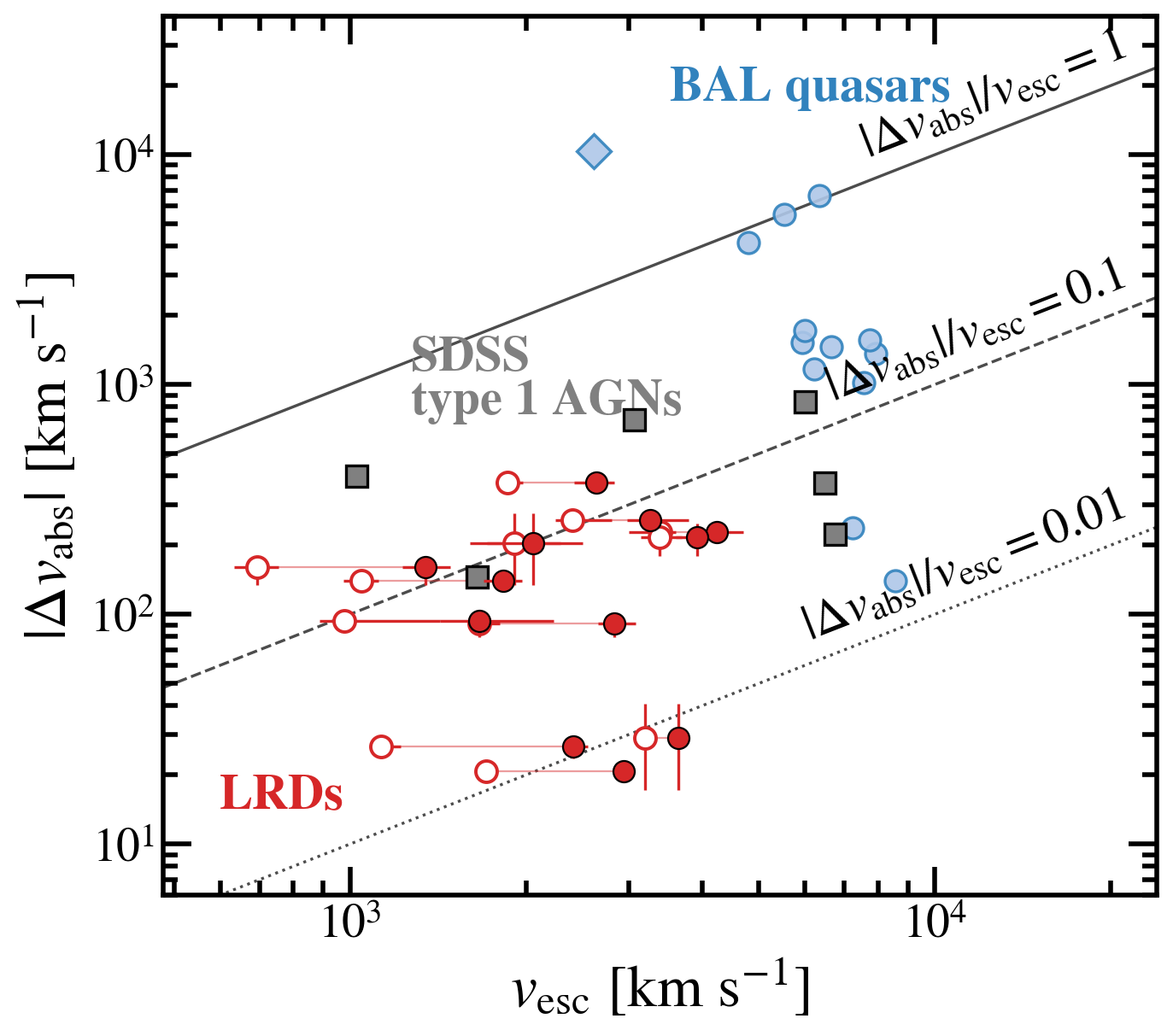}
    \caption{Absolute velocity offset of the H$\alpha$ absorbers, $|\dvabs(\ha)|$, as a function of the escape velocity at the BLR radius, $v_{\rm esc}=\sqrt{2f}\,\mathrm{FWHM}$ (Equation~\ref{eq:vesc}). Only blueshifted (outflowing) absorbers are shown. Red filled circles show the LRD absorbers with $v_{\rm esc}$ computed from the observed FWHM including the electron-scattering wings, while red open circles show the same objects with $v_{\rm esc}$ from the intrinsic Gaussian FWHM corrected for electron scattering (each pair is connected by a thin line). Gray squares show the six blueshifted absorption components of low-$z$ SDSS type 1 AGNs from \citet{Shangguan+2026}. Blue circles show the 12 blueshifted FeLoBAL quasars with Balmer absorption from \citet{Leighly+2025}, and the blue diamond shows SDSS J1523$+$3914 \citep{Zhang+2015}. Because \citet{Leighly+2025} do not publish broad-line FWHMs, their $v_{\rm esc}$ is derived from the published $M_{\rm BH}$ using $L_{\rm bol}/L_{5100}=9$ and the radius--luminosity relation of \citet{Bentz+2013}. Diagonal lines indicate constant $|\dvabs(\ha)|/v_{\rm esc}=1$, 0.1, and 0.01.}
    \label{fig:dv_vesc}
\end{figure}

To normalize the absorber velocities, we evaluate the escape velocity at the radius of the broad-line region (BLR). For a virial black hole mass $M_\mathrm{BH}=f\,R_{\rm BLR}\,\mathrm{FWHM^2_{H\alpha,broad}}/G$, this velocity is
\begin{equation}\label{eq:vesc}
v_{\rm esc}=\sqrt{\frac{2GM_\mathrm{BH}}{R_{\rm BLR}}}=\sqrt{2f}\,\mathrm{FWHM_{H\alpha,broad}},
\end{equation}
Thus, $\dvabs(\ha)/v_{\rm esc}$ is independent of the BLR radius--luminosity relation, with the virial factor entering only as $\sqrt{f}$; we adopt $f=1.12$ \citep{Woo+2015}. For our sample, we use the FWHM of the total broad-H$\alpha$ model profile including the electron-scattering wings (Section~\ref{sec:line_profile_fitting}), and also examine the intrinsic Gaussian widths corrected for electron scattering \citep[cf.][]{Rusakov+2026}. For the SDSS absorbers, we use the broad-H$\alpha$ widths reported by \citet{Shangguan+2026}. Figure~\ref{fig:dv_vesc} shows the absorbers in the $|\dvabs(\ha)|$--$v_{\rm esc}$ plane, together with lines of constant $|\dvabs(\ha)|/v_{\rm esc}$.

Because the escape analysis below concerns outward-moving gas, we restrict this comparison to the blueshifted absorbers, namely 12 of our 14 LRD absorbers and the six blueshifted SDSS components. The median normalized offset is $|\dvabs(\ha)|/v_{\rm esc}\simeq0.056$ for the LRD absorbers and $\simeq0.115$ for the SDSS absorption components. The difference is not significant according to either the Anderson--Darling ($p=0.057$) or the Kolmogorov--Smirnov test ($p=0.25$), whereas the unnormalized offsets differ (Anderson--Darling $p=0.012$). If we instead use the intrinsic Gaussian widths for the LRD sample, the median becomes $|\dvabs(\ha)|/v_{\rm esc}\simeq0.081$, and the two distributions are statistically indistinguishable (Anderson--Darling $p>0.25$; Kolmogorov--Smirnov $p=0.76$).
Thus, the factor of $\sim2.6$ difference in the absolute velocity offsets (medians of 151 and $387\ {\rm km\,s^{-1}}$) is substantially reduced after normalization by the BLR dynamical velocity scale. This is consistent with the lower absolute velocities of LRDs partly reflecting their smaller characteristic nuclear velocities, possibly due to the shallower gravitational potential. 

Both the LRD and SDSS Balmer absorbers occupy a low-$|\dvabs(\ha)|/v_{\rm esc}$ regime, with $|\dvabs(\ha)|/v_{\rm esc}\lesssim0.4$. For the LRDs, the maximum ratio is 0.142 for the observed broad-line widths and 0.232 for the intrinsic widths. Balmer absorbers in FeLoBAL quasars extend to considerably higher values: the 12 blueshifted systems of \citet{Leighly+2025} span $|\dvabs(\ha)|=140$--$6620\ {\rm km\,s^{-1}}$ and $|\dvabs(\ha)|/v_{\rm esc}\simeq0.02$--1.0, with a median of 0.21. Most of them nevertheless belong to the phenomenological low-velocity ``loitering'' class with $|\dvabs(\ha)|<2000\ {\rm km\,s^{-1}}$, and only SDSS J1625$+$0933 reaches $|\dvabs(\ha)|/v_{\rm esc}>1$. The extreme Balmer BAL quasar SDSS J1523$+$3914, which is not covered by that sample, lies far above at $|\dvabs(\ha)|/v_{\rm esc}\simeq4$ \citep{Zhang+2015}. 

\subsection{Conditional Escape in the Presence of Radiation Pressure}\label{sec:radiation}
Whether an absorber ultimately escapes is a separate question from the kinematic scaling above. The absorbing clouds are exposed to radiation from the central engine, so gravity and radiation pressure must be included in the same equation of motion. In the idealized limit adopted here, the incident radiation transfers its full single-scattering momentum through the illuminated surface of a fixed-column cloud, while the cloud inertia grows in proportion to its column density. This complete-coupling assumption maximizes the radiative acceleration at fixed bolometric luminosity. The outward radiative acceleration relative to gravity is then quantified by
\begin{equation}\label{eq:gammaeff}
\Gamma_{\rm eff}\simeq\frac{\lambda_{\rm Edd}}{\sigma_{\rm T}N_{\rm H}},
\end{equation}
where $\lambda_{\rm Edd}=L/L_{\rm Edd}$ is the Eddington ratio and $\sigma_\mathrm{T}$ is the Thomson scattering cross section. Because radiation pressure and gravity both scale as $r^{-2}$ in this approximation, a radial cloud of fixed column density obeys $\ddot{r}=-GM_\mathrm{BH}(1-\Gamma_{\rm eff})/r^{2}$. Integrating the equation of motion, an absorber observed at $R_{\rm BLR}$ has non-negative specific energy (and, if moving outward, can reach infinity) when
\begin{equation}\label{eq:escape}
\mathcal{E}\;\equiv\;\left(\frac{\dvabs(\ha)}{v_{\rm esc}}\right)^{2}+\Gamma_{\rm eff}\;>\;1 .
\end{equation}
This criterion is conditional on radial motion, a point-mass potential, a fixed column density, and sustained illumination; the observed line-of-sight centroid is used as a proxy for the radial speed. The resulting plane is analogous to the ``forbidden region'' formalism for dusty AGN obscurers \citep[e.g.,][]{Fabian+2008, Ricci+2017}, but here adopts the complete single-scattering cloud approximation above and retains the measured kinematic term explicitly.

\begin{figure}
    \centering
    \includegraphics[width=1\linewidth]{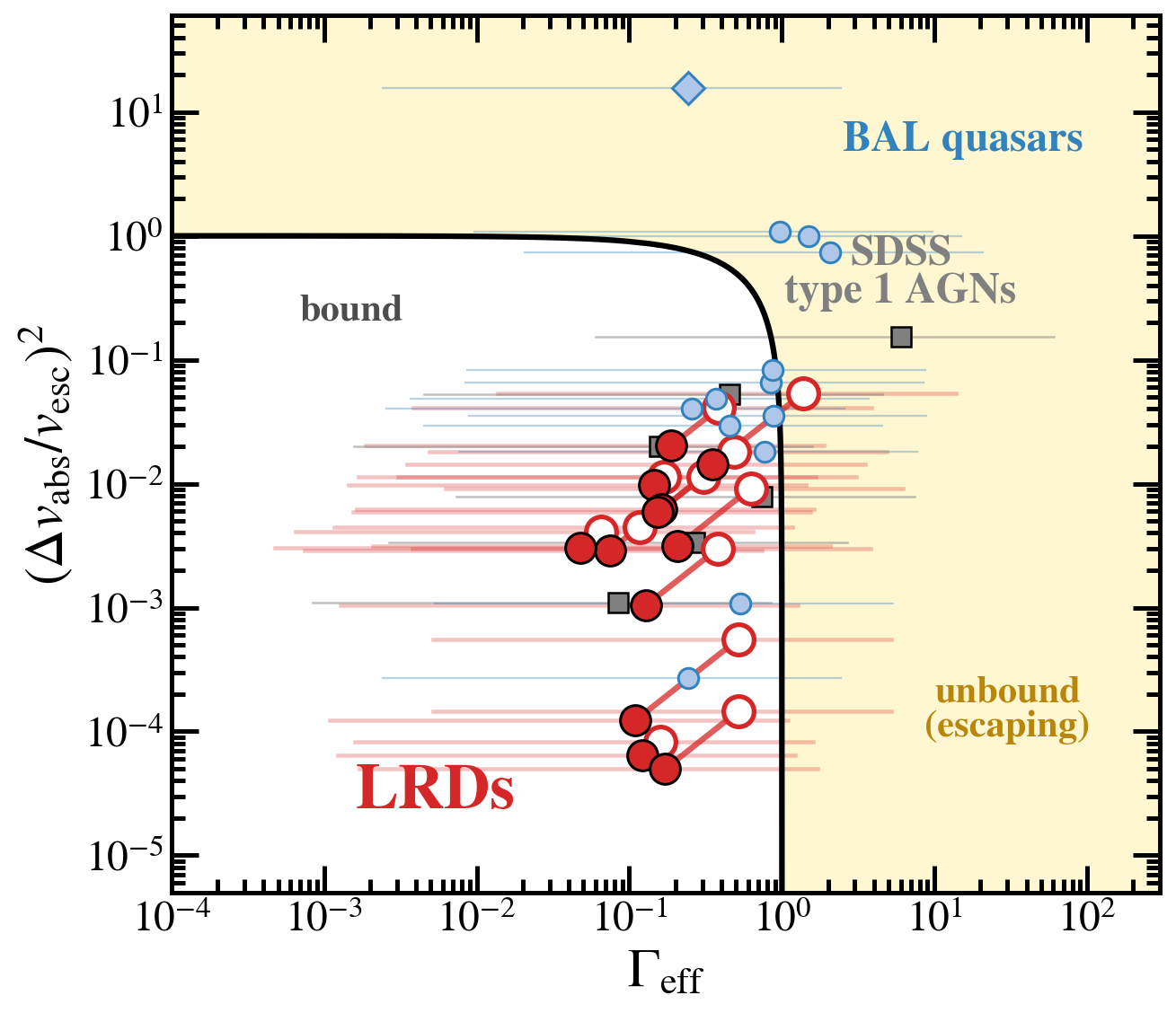}
    \caption{Escape diagram based on the conditional criterion of Equation~(\ref{eq:escape}). Only blueshifted (outflowing) absorbers are shown. The black curve marks $\mathcal{E}=1$, and the yellow shaded region is unbound within the idealized fixed-column, radial, point-mass model. Symbols are the same as Figure \ref{fig:dv_vesc}. Symbols are placed at a fiducial $N_{\rm H}=10^{24}\ {\rm cm^{-2}}$; the horizontal spans trace the assumed range $N_{\rm H}=10^{23}$--$10^{26}\ {\rm cm^{-2}}$ and are model tracks rather than measurement uncertainties.
    }
    \label{fig:escape}
\end{figure}

For absorbers with the kinematic term $(\dvabs(\ha)/v_{\rm esc})^{2}$ below unity, solving $\mathcal{E}=1$ for the column density yields the critical value below which an absorber is energetically unbound,
\begin{equation}\label{eq:nhcrit}
N_{\rm H,crit}=\frac{\lambda_{\rm Edd}}{\sigma_{\rm T}\left[1-(\dvabs(\ha)/v_{\rm esc})^{2}\right]}.
\end{equation}
For the LRD points in Figure~\ref{fig:escape}, which use the observed-profile FWHMs, $(|\dvabs(\ha)|/v_{\rm esc})^{2}\lesssim0.02$, so $N_{\rm H,crit}\simeq1.5\times10^{24}\,\lambda_{\rm Edd}\ {\rm cm^{-2}}$; the velocity term changes this threshold by only a few per cent. Estimating $\lambda_{\rm Edd}$ for each LRD from $M_\mathrm{BH}$ \citep{Greene_Ho_2005} and the LRD-specific bolometric correction $L_{\rm bol}=19\,\lbha$ \citep{Greene+2025}, we obtain a median $\lambda_{\rm Edd}\simeq0.10$ and a median $\log(N_{\rm H,crit}/{\rm cm^{-2}})\simeq23.2$ (range 22.7--23.6) for the observed widths, and $\lambda_{\rm Edd}\simeq0.25$ with $\log(N_{\rm H,crit}/{\rm cm^{-2}})\simeq23.6$ (range 22.8--24.2) if the intrinsic widths are adopted instead. %Because complete coupling was assumed, these critical columns are upper limits if only a fraction of $L_{\rm bol}$ couples to the cloud.
At the fiducial $N_{\rm H}=10^{24}\ {\rm cm^{-2}}$, the LRD absorbers therefore lie predominantly on the bound side of the conditional criterion. By contrast, adopting $N_{\rm H}\sim10^{23}\ {\rm cm^{-2}}$ moves most of them to the unbound side. Their model tracks thus cross the escape boundary over a physically plausible (but presently unconstrained) range of total column densities. The $10^{24}$--$10^{26}\ {\rm cm^{-2}}$ columns invoked by dense-envelope models \citep[e.g.,][]{Ji+2025, Naidu+2025} favor the bound branch, whereas the columns inferred from the Balmer break and decrement \citep[$\log N_{\rm H}\simeq22.4$--23.8;][]{Chen+2026} and from X-ray spectral fitting of the two X-ray-detected LRDs \citep[$\log N_{\rm H}\simeq22.7$ and 23.3;][]{Kocevski+2025} overlap the transition. Figure~\ref{fig:escape} therefore does not establish that any individual LRD absorber is bound, unbound, or fine-tuned to the boundary; rather, it identifies the range of $N_{\rm H}$ over which its inferred fate changes. The FeLoBAL quasars of \citet{Leighly+2025} reach larger kinematic terms than the LRDs (median $(\dvabs(\ha)/v_{\rm esc})^{2}\simeq0.04$), but only one of the 12 exceeds $\mathcal{E}=1$ through the kinematic term alone; their fate is likewise governed mainly by $\Gamma_{\rm eff}$.

%Equation~(\ref{eq:escape}) also parameterizes the unknown absorber radius: for fixed $\Gamma_{\rm eff}<1$, $r_{\rm crit}/R_{\rm BLR}=(1-\Gamma_{\rm eff})/(|\dvabs(\ha)|/v_{\rm esc})^{2}$. However, neither $r$ nor total $N_{\rm H}$ is currently measured for the LRD absorbers. The high density required to maintain a substantial $n=2$ hydrogen population may favor a nuclear location \citep[e.g.,][]{Inayoshi_Maiolino_2025}, but it does not determine the absorber radius relative to $R_{\rm BLR}$.
The fact that the inferred critical columns fall within the broad range of columns discussed for LRDs suggests that Balmer absorbers may probe the transition between escaping and failed or recycling flows.  %without implying that every system is observed exactly at this boundary. 
On the low-$\Gamma_\mathrm{eff}$ side, the observed blueshift indicates current outward motion, but $\mathcal{E}<1$ implies that the cloud cannot escape in the adopted model: it will stall and eventually fall back. Such a trajectory is analogous to failed-wind calculations in which radiatively lifted BLR clouds lose their radiative support and return toward the disk \citep[e.g.,][]{Czerny_Hryniewicz_2011, Owen_Lin_2025}. Near its turning point, the cloud could appear as a low-velocity ``loitering'' or quasi-static component, qualitatively reminiscent of compact low-velocity FeLoBAL absorbers \citep{Leighly+2025} and of the nearly systemic Balmer absorption attributed to a dense shell or atmosphere in an extreme LRD \citep{Naidu+2025}. On the high-$\Gamma_\mathrm{eff}$ side, radiation pressure can instead unbind and expel the cloud. 
%If the Balmer absorber is part of the same high-covering, optically thick medium that produces the LRD spectral appearance, such clearing could reduce the covering of the central engine and contribute to the end of the LRD phase. This evolutionary interpretation remains speculative because a line-of-sight Balmer absorber need not represent the global obscuring medium. 
The fate of the gas is controlled mainly by the presently uncertain ratio $\Gamma_\mathrm{eff}\propto\lambda_{\rm Edd}/N_{\rm H}$; direct constraints on the Eddington ratio and total column density are required to distinguish radiatively escaping gas from a failed flow. 

We note that this model treats $N_{\rm H}$ and $\lambda_{\rm Edd}$ as constant in time, so a bound orbit is time-reversal symmetric and would spend equal time moving outward and inward -- in tension with the observed excess of blueshifted absorbers. This tension suggests that clouds may become more prone to escape as they move outward (e.g., if their column density decreases with time), but capturing such time evolution would require time-dependent radiation-hydrodynamic modeling, which is beyond the scope of this work.

\section{Summary}\label{sec:summary}
In this work, we use archival JWST/NIRSpec spectra to present a statistical analysis of Balmer absorption in a color-selected sample of 40 LRDs. Our results are summarized below:

\begin{itemize}
    \item We construct a sample of 40 color-selected LRDs from the DJA NIRSpec data set and complementary NIRSpec/IFU observations. We fit the H$\alpha$+[N\,\textsc{ii}] line profiles with narrow and broad emission components, allowing for a partial-covering Balmer absorption component parameterized by optical depth and covering factor when required by the data.

    \item We find that Balmer absorption is much more frequently detected in LRDs than in low-redshift type 1 AGNs: the incidence is $14/40=35\%$ in our LRD sample, compared with $6/14583\simeq0.04\%$ in SDSS type 1 AGNs. Seven of the 14 LRD absorbers are detected with medium-resolution gratings, so the approximately 850-fold difference cannot be explained solely by the spectral resolutions of NIRSpec and SDSS. This high incidence suggests that dense gas capable of producing Balmer absorption has a larger line-of-sight covering probability in LRDs than in ordinary type 1 AGNs. Because fully synthetic tests by \citet{Juodzbalis+2026} indicate substantial incompleteness, particularly near the systemic velocity and at $R\sim1000$, our measured LRD incidence should be regarded as a conservative lower limit.

    \item We find that Balmer absorption in LRDs is predominantly blueshifted relative to the systemic redshift traced by [O\,\textsc{iii}], with $12/14$ absorbers showing $\dvabs(\ha)<0$. This indicates that the detected Balmer absorbers in LRDs are preferentially associated with outflowing gas along the line of sight.

    \item Compared with the SDSS type 1 AGNs with Balmer absorption from \citet{Shangguan+2026}, the LRD absorbers tend to have smaller velocity offsets. A forward model that injects empirical SDSS absorption profiles into semi-synthetic LRD spectra shows that the object-dependent selection function does not reproduce the narrow observed distribution. 
    
    \item Normalizing the velocity offsets of the blueshifted absorbers by the BLR escape velocity reduces the difference in the medians to a factor of $\sim1.4$--2.1. This is consistent with the lower absolute velocities of LRDs partly tracking their smaller nuclear velocity scales. In the simple analytic model including gravity and radiative force, at the fiducial $N_{\rm H}=10^{24}\ {\rm cm^{-2}}$, the LRD absorbers lie predominantly on the bound side of the conditional criterion, whereas $N_{\rm H}\sim10^{23}\ {\rm cm^{-2}}$ places most of them on the unbound side. The plausible column range therefore straddles a transition between escaping outflows and failed or recycling flows: high-column clouds may stall and fall back, while low-column clouds can be expelled. 
\end{itemize}

\begin{acknowledgments}
We thank Harley Katz, Marta Volonteri, Xiaojing Lin, Jorryt Matthee, Xihan Ji, and Roberto Maiolino for valuable discussions. 
This work is based on observations made with the NASA/ ESA/CSA James Webb Space Telescope. The NIRSpec/IFU data were obtained from the Mikulski Archive for Space Telescopes at the Space Telescope Science Institute, which is operated by the Association of Universities for Research in Astronomy, Inc., under NASA contract NAS 5-03127 for JWST. The NIRSpec MSA data presented herein were retrieved from DJA. DJA is an initiative of the Cosmic Dawn Center (DAWN), which is funded by the Danish National Research Foundation under grant DNRF140. We thank DAWN for providing the reduced NIRSpec data. The observational data presented in this work are associated with programs 
ERS 1345 (CEERS; PI: Finkelstein), GTO 1180 (PI: Eisenstein), GTO 1181 (PI: Eisenstein), GTO 1210 (PI: Luetzgendorf), GTO 1286 (PI: Luetzgendorf), GO 1879 (PI: Curti), GO 2674 (PI: Arrabal Haro), GO 3567 (DeepDive; PI: Valentino), GO 4106 (PI: Nelson), GO 4233 (PI: de Graaff), GO 4750 (PI: Nakajima), GO 4762 (PI: Fujimoto), GO 5015 (PI: Übler and Maiolino), GO 5664 (PI: Matthee), and DD 9223 (PI: Fujimoto). The authors acknowledge the teams conducting these observations for publicly releasing the data. This work benefited from discussions at the workshop “Continuing the JWST Revolution: Understanding Early Galaxy Formation,” hosted by the Munich Institute for Astro-, Particle and BioPhysics (MIAPbP). 
HY acknowledges support by KAKENHI (25KJ0832) through Japan Society for the Promotion of Science (JSPS). 
MO acknowledges the supports from the World Premier International Research Center Initiative (WPI Initiative), MEXT, Japan, the joint research program of the Institute for Cosmic Ray Research (ICRR), the University of Tokyo, and KAKENHI (21H04467, 25H00674) through Japan Society for the Promotion of Science (JSPS). YK acknowledges support by KAKENHI (26KJ0960) through JSPS, JSR Fellowship, and FoPM, WINGS Program, the University of Tokyo. TK acknowledges support by KAKENHI (26KJ1232) through Japan Society for the Promotion of Science (JSPS). MN is supported by JSPS KAKENHI Grant Nos. 25KJ0828. YH acknowledges support from the Japan Society for the Promotion of Science (JSPS) Grant-in-Aid for Scientific Research (24H00245) and the JSPS International Leading Research (22K21349). 
The authors acknowledge the use of ChatGPT (OpenAI, GPT-5.6), Codex (OpenAI, Codex 5.6), and Claude (Anthropic, Sonnet 5 and Fable 5) to assist with language editing, code development, and figure generation. All AI-assisted outputs were carefully reviewed and validated by the authors. The authors take full responsibility for all analyses, interpretations, and conclusions presented in this work.

\end{acknowledgments}

\facilities{JWST}

\software{
grizli \citep{Brammer_2021, Brammer_2023}, emcee \citep{Foreman-Mackey+13}, 
Matplotlib \citep{Hunter_2007}, Astropy \citep{Astropy+2013,Astropy+2018,Astropy+2022}, NumPy \citep{Harris+2020} 
          }

\clearpage
\appendix
\section{Compiled Balmer-Absorber Properties}\label{app:compiled_balmer_properties}

\startlongtable
\begin{deluxetable*}{lcccccccc}
\tabletypesize{\scriptsize}
\tablecaption{Reported H$\alpha$ Emission Properties of Balmer Absorbers\label{tab:balmer_emission_properties}}
\tablewidth{0pt}
\tablehead{
\colhead{ID} & \colhead{R.A.} & \colhead{Decl.} & \colhead{$z$} & \colhead{$\log L_{\mathrm{H}\alpha,\mathrm{b}}$} & \colhead{$\mathrm{EW}_{\mathrm{H}\alpha,\mathrm{b}}$} & \colhead{$\mathrm{FWHM}_{\mathrm{H}\alpha,\mathrm{b}}^{\mathrm{obs}}$} & \colhead{$\mathrm{FWHM}_{\mathrm{H}\alpha,\mathrm{b}}^{\mathrm{int}}$} & \colhead{Ref.} \\
\colhead{} & \colhead{deg} & \colhead{deg} & \colhead{} & \colhead{erg s$^{-1}$} & \colhead{\AA} & \colhead{km s$^{-1}$} & \colhead{km s$^{-1}$} & \colhead{}
}
\startdata
EGS-47962 & 214.89248 & 52.85689 & 6.72701 & $42.864_{-0.051}^{+0.062}$ & $407_{-87}^{+105}$ & $1378_{-238}^{+281}$ & $1278_{-206}^{+254}$ & this work \\
JADES-GN-68797 & 189.22914 & 62.14619 & 5.03923 & $43.660_{-0.013}^{+0.049}$ & $715_{-66}^{+232}$ & $2838_{-272}^{+296}$ & $2270_{-267}^{+137}$ & this work \\
 &  &  &  & $43.736_{-0.001}^{+0.001}$ & $\ldots$ & $2009_{-11}^{+9}$ & $\ldots$ & IJ26 \\
 &  &  &  & $\ldots$ & $\ldots$ & $1414\pm22$ & $\ldots$ & JM26 \\
 &  &  &  & $\ldots$ & $\ldots$ & $\ldots$ & $\ldots$ & CC26 \\
RUBIES-EGS-49140 & 214.89225 & 52.87741 & 6.68494 & $43.842_{-0.012}^{+0.013}$ & $1217_{-106}^{+206}$ & $2436_{-88}^{+94}$ & $2135_{-63}^{+66}$ & this work \\
 &  &  &  & $43.781_{-0.006}^{+0.006}$ & $\ldots$ & $2475_{-45}^{+46}$ & $\ldots$ & IJ26 \\
 &  &  &  & $\ldots$ & $\ldots$ & $1446\pm43$ & $\ldots$ & JM26 \\
 &  &  &  & $\ldots$ & $\ldots$ & $\ldots$ & $\ldots$ & KD26 \\
 &  &  &  & $\ldots$ & $\ldots$ & $\ldots$ & $\ldots$ & CC26 \\
 &  &  &  & $\ldots$ & $\ldots$ & $2613_{-39}^{+48}$ & $\ldots$ & AT25 \\
 &  &  &  & $43.619\pm0.002$ & $6638_{-232}^{+234}$ & $2909\pm31$ & $\ldots$ & EL25 \\
 &  &  &  & $44.19\pm0.05$ & $\ldots$ & $2220_{-10}^{+20}$ & $1350\pm50$ & FD25 \\
RUBIES-EGS-42046 & 214.79537 & 52.78885 & 5.27623 & $43.614_{-0.011}^{+0.013}$ & $1062_{-155}^{+188}$ & $1760_{-146}^{+139}$ & $1242_{-56}^{+78}$ & this work \\
 &  &  &  & $43.687_{-0.006}^{+0.006}$ & $\ldots$ & $2003_{-53}^{+51}$ & $\ldots$ & IJ26 \\
 &  &  &  & $\ldots$ & $\ldots$ & $1430\pm40$ & $\ldots$ & JM26 \\
 &  &  &  & $\ldots$ & $\ldots$ & $\ldots$ & $\ldots$ & KD26 \\
 &  &  &  & $\ldots$ & $\ldots$ & $\ldots$ & $\ldots$ & CC26 \\
 &  &  &  & $\ldots$ & $\ldots$ & $3146_{-67}^{+75}$ & $\ldots$ & AT25 \\
RUBIES-EGS-55604 & 214.98303 & 52.95600 & 6.98354 & $43.899_{-0.017}^{+0.019}$ & $1506_{-210}^{+550}$ & $2218_{-245}^{+328}$ & $1604_{-104}^{+269}$ & this work \\
 &  &  &  & $43.636_{-0.005}^{+0.004}$ & $\ldots$ & $1922_{-38}^{+41}$ & $\ldots$ & IJ26 \\
 &  &  &  & $\ldots$ & $\ldots$ & $1443\pm52$ & $\ldots$ & JM26 \\
 &  &  &  & $\ldots$ & $\ldots$ & $\ldots$ & $\ldots$ & CC26 \\
DeepDive-20910 & 214.94902 & 52.85175 & 3.71973 & $42.917_{-0.014}^{+0.018}$ & $709_{-55}^{+107}$ & $1907_{-108}^{+127}$ & $1500_{-57}^{+94}$ & this work \\
DeepDive-51909 & 34.45493 & -5.13271 & 4.62577 & $43.382_{-0.010}^{+0.011}$ & $1293_{-96}^{+130}$ & $1911_{-134}^{+147}$ & $1113_{-49}^{+92}$ & this work \\
GN-9771 & 189.28100 & 62.24730 & 5.53457 & $43.735_{-0.002}^{+0.002}$ & $1618_{-50}^{+27}$ & $1960_{-16}^{+16}$ & $1144_{-5}^{+5}$ & this work \\
 &  &  &  & $43.740_{-0.022}^{+0.021}$ & $\ldots$ & $2312_{-299}^{+299}$ & $\ldots$ & IJ26 \\
 &  &  &  & $\ldots$ & $\ldots$ & $1528\pm9$ & $\ldots$ & JM26 \\
 &  &  &  & $43.650_{-0.011}^{+0.011}$ & $\ldots$ & $3739\pm112$ & $\ldots$ & JM24 \\
 &  &  &  & $\ldots$ & $\ldots$ & $1549\pm5$ & $\ldots$ & AT26a \\
BT-159717 & 53.09750 & -27.90130 & 5.07746 & $42.906_{-0.005}^{+0.005}$ & $866_{-29}^{+29}$ & $1618_{-77}^{+82}$ & $754_{-31}^{+61}$ & this work \\
 &  &  &  & $42.934_{-0.012}^{+0.011}$ & $\ldots$ & $1617_{-67}^{+67}$ & $\ldots$ & IJ26 \\
 &  &  &  & $\ldots$ & $\sim1100$ & $1490\pm80$ & $\ldots$ & FD26b \\
UNCOVER-4286 & 3.61920 & -30.42330 & 5.83150 & $42.798_{-0.016}^{+0.016}$ & $622_{-56}^{+69}$ & $1156_{-209}^{+328}$ & $653_{-61}^{+299}$ & this work \\
GN-15498 & 189.28550 & 62.28080 & 5.08465 & $42.694_{-0.009}^{+0.008}$ & $542_{-41}^{+27}$ & $1222_{-96}^{+95}$ & $698_{-48}^{+50}$ & this work \\
 &  &  &  & $42.723_{-0.003}^{+0.003}$ & $\ldots$ & $1162_{-10}^{+10}$ & $\ldots$ & IJ26 \\
 &  &  &  & $\ldots$ & $\ldots$ & $1048\pm9$ & $\ldots$ & JM26 \\
GS-13971 & 53.13860 & -27.79030 & 5.48161 & $42.882_{-0.005}^{+0.005}$ & $1006_{-50}^{+62}$ & $903_{-88}^{+87}$ & $464_{-40}^{+41}$ & this work \\
 &  &  &  & $42.935_{-0.041}^{+0.038}$ & $\ldots$ & $1384_{-185}^{+185}$ & $\ldots$ & IJ26 \\
 &  &  &  & $\ldots$ & $\ldots$ & $1083\pm35$ & $\ldots$ & JM26 \\
RUBIES-182791 & 34.21381 & -5.08705 & 4.71492 & $43.083_{-0.032}^{+0.037}$ & $1215_{-255}^{+490}$ & $2622_{-217}^{+260}$ & $2260_{-159}^{+208}$ & this work \\
 &  &  &  & $42.869_{-0.008}^{+0.009}$ & $\ldots$ & $1752_{-96}^{+87}$ & $\ldots$ & IJ26 \\
 &  &  &  & $\ldots$ & $\ldots$ & $1602\pm50$ & $\ldots$ & JM26 \\
BT-J1148 & 177.05796 & 52.86280 & 5.01148 & $42.840_{-0.017}^{+0.012}$ & $409_{-34}^{+48}$ & $1441_{-62}^{+65}$ & $1213_{-36}^{+37}$ & this work \\
 &  &  &  & $42.734_{-0.002}^{+0.002}$ & $\ldots$ & $956_{-6}^{+6}$ & $\ldots$ & IJ26 \\
 &  &  &  & $42.519_{-0.104}^{+0.084}$ & $\ldots$ & $2886\pm346$ & $\ldots$ & JM24 \\
JADES-GN-28074 & 189.06458 & 62.27382 & 2.260 & $43.400_{-0.001}^{+0.001}$ & $\ldots$ & $1750_{-6}^{+5}$ & $\ldots$ & IJ26 \\
 &  &  &  & $\ldots$ & $\ldots$ & $\ldots$ & $\ldots$ & CC26 \\
 &  &  &  & $43.083_{-0.004}^{+0.004}$ & $\ldots$ & $3610_{-22}^{+21}$ & $\ldots$ & IJ24 \\
DH-GS-5070 & 53.09200 & -27.90314 & 3.642 & $42.745_{-0.002}^{+0.002}$ & $\ldots$ & $1785_{-14}^{+14}$ & $\ldots$ & IJ26 \\
JADES-GN-38147 & 189.27068 & 62.14842 & 5.869 & $42.996_{-0.008}^{+0.009}$ & $\ldots$ & $1675_{-49}^{+55}$ & $\ldots$ & IJ26 \\
 &  &  &  & $\ldots$ & $\ldots$ & $1040\pm63$ & $\ldots$ & JM26 \\
 &  &  &  & $\ldots$ & $\ldots$ & $\ldots$ & $\ldots$ & CC26 \\
 &  &  &  & $\ldots$ & $\ldots$ & $920\pm110$ & $480\pm50$ & LI26 \\
The Cliff & 34.41075 & -5.12966 & 3.549 & $42.602_{-0.002}^{+0.001}$ & $\ldots$ & $865_{-7}^{+2}$ & $\ldots$ & IJ26 \\
 &  &  &  & $\ldots$ & $\ldots$ & $1400\pm12$ & $\ldots$ & JM26 \\
 &  &  &  & $\ldots$ & $\sim400$ & $1533_{-80}^{+110}$ & $\ldots$ & AdG25 \\
A2744-QSO1 & 3.57984 & -30.40157 & 7.036 & $42.173_{-0.006}^{+0.055}$ & $\ldots$ & $816_{-10}^{+11}$ & $\ldots$ & IJ26 \\
 &  &  &  & $\ldots$ & $170_{-10}^{+20}$ & $680_{-80}^{+70}$ & $\ldots$ & FD26a \\
A2744-45924 & 3.58476 & -30.34363 & 4.464 & $\ldots$ & $\ldots$ & $1485\pm18$ & $\ldots$ & JM26 \\
 &  &  &  & $\sim44.0$ & $>800$ & $4540\pm50$ & $\ldots$ & IL24 \\
OCEANS-102364 & 215.02207 & 52.92079 & 4.542 & $\ldots$ & $\ldots$ & $\ldots$ & $\ldots$ & KD26 \\
OCEANS-100424 & 214.88680 & 52.85538 & 4.953 & $\ldots$ & $\ldots$ & $\ldots$ & $\ldots$ & KD26 \\
RUBIES-UDS-146995 & 34.33104 & -5.13996 & 3.732 & $\ldots$ & $\ldots$ & $1403_{-89}^{+85}$ & $\ldots$ & AT25 \\
RUBIES-EGS-28812 & 214.92415 & 52.84905 & 4.222 & $\ldots$ & $\ldots$ & $2098_{-117}^{+120}$ & $\ldots$ & AT25 \\
PAN-BH*-1 & 40.01584 & -1.65936 & 1.731 & $\ldots$ & $\ldots$ & $1257\pm27$ & $\ldots$ & AT26b \\
J102530.29+140207.3 (Egg) & 156.37621 & 14.03536 & 0.1007 & $41.914_{-0.013}^{+0.013}$ & $\ldots$ & $956\pm27$ & $\ldots$ & XL25 \\
 &  &  &  & $41.903_{-0.002}^{+0.002}$ & $\ldots$ & $934\pm10$ & $520\pm10$ & XJ26 \\
J102208.52+084156.1 & 155.53550 & 8.69892 & 0.2227 & $42.716_{-0.011}^{+0.011}$ & $\ldots$ & $805\pm34$ & $\ldots$ & XL25 \\
J012930.87+062843.32 & 22.37862 & 6.47870 & 0.2467 & $42.142_{-0.005}^{+0.005}$ & $\ldots$ & $744_{-27}^{+27}$ & $\ldots$ & XL26 \\
 &  &  &  & $\ldots$ & $\ldots$ & $748\pm9$ & $\ldots$ & KP26 \\
J082606.37-010001.31 & 126.52654 & -1.00036 & 0.6273 & $42.935_{-0.035}^{+0.067}$ & $\ldots$ & $677_{-152}^{+128}$ & $\ldots$ & XL26 \\
J082921.37+131237.44 & 127.33904 & 13.21040 & 0.3986 & $42.647_{-0.035}^{+0.057}$ & $\ldots$ & $691_{-97}^{+101}$ & $\ldots$ & XL26 \\
J094411.31-024908.65 & 146.04712 & -2.81907 & 0.6623 & $43.513_{-0.017}^{+0.012}$ & $\ldots$ & $552_{-20}^{+22}$ & $\ldots$ & XL26 \\
J102553.75+502843.24 & 156.47396 & 50.47868 & 0.8824 & $43.767_{-0.006}^{+0.006}$ & $\ldots$ & $699_{-41}^{+36}$ & $\ldots$ & XL26 \\
J132137.00-021417.04 & 200.40417 & -2.23807 & 0.2244 & $41.613_{-0.006}^{+0.006}$ & $\ldots$ & $272_{-6}^{+6}$ & $\ldots$ & XL26 \\
J142337.59+520216.05 & 215.90662 & 52.03779 & 0.6236 & $42.900_{-0.016}^{+0.012}$ & $\ldots$ & $1248_{-65}^{+88}$ & $\ldots$ & XL26 \\
J161102.44+091728.60 & 242.76017 & 9.29128 & 0.6952 & $43.515_{-0.212}^{+0.018}$ & $\ldots$ & $811_{-33}^{+952}$ & $\ldots$ & XL26 \\
J164102.65+070806.47 & 250.26104 & 7.13513 & 0.5351 & $42.803_{-0.019}^{+0.022}$ & $\ldots$ & $775_{-82}^{+72}$ & $\ldots$ & XL26 \\
J164637.91+142648.62 & 251.65796 & 14.44684 & 0.7071 & $43.446_{-0.029}^{+0.018}$ & $\ldots$ & $878_{-55}^{+100}$ & $\ldots$ & XL26 \\
J165450.36+033741.74 & 253.70983 & 3.62826 & 0.6408 & $43.238_{-0.018}^{+0.018}$ & $\ldots$ & $1082_{-52}^{+41}$ & $\ldots$ & XL26 \\
J171741.74+380752.47 & 259.42392 & 38.13124 & 0.1959 & $42.655_{-0.014}^{+0.015}$ & $\ldots$ & $960_{-134}^{+201}$ & $\ldots$ & XL26 \\
 &  &  &  & $\ldots$ & $\ldots$ & $1248\pm10$ & $\ldots$ & KP26 \\
J212725.88-044808.92 & 321.85783 & -4.80248 & 0.5842 & $42.686_{-0.033}^{+0.042}$ & $\ldots$ & $921_{-152}^{+140}$ & $\ldots$ & XL26 \\
J225535.58+154216.29 & 343.89825 & 15.70453 & 0.4273 & $42.061_{-0.034}^{+0.040}$ & $\ldots$ & $814_{-110}^{+124}$ & $\ldots$ & XL26 \\
J111943.20+021911.32 & 169.93000 & 2.31981 & 0.4682 & $41.975_{-0.019}^{+0.019}$ & $\ldots$ & $742_{-112}^{+151}$ & $\ldots$ & XL26 \\
J113734.35+552028.16 & 174.39312 & 55.34116 & 0.4358 & $42.092_{-0.014}^{+0.016}$ & $\ldots$ & $593_{-53}^{+60}$ & $\ldots$ & XL26 \\
 &  &  &  & $\ldots$ & $\ldots$ & $891\pm39$ & $\ldots$ & KP26 \\
J134317.81+393418.07 & 205.82421 & 39.57169 & 0.2933 & $41.822_{-0.009}^{+0.010}$ & $\ldots$ & $556_{-21}^{+22}$ & $\ldots$ & XL26 \\
 &  &  &  & $\ldots$ & $\ldots$ & $588\pm16$ & $\ldots$ & KP26 \\
J190954.15+583112.37 & 287.47562 & 58.52010 & 0.4273 & $42.147_{-0.094}^{+0.237}$ & $\ldots$ & $483_{-111}^{+172}$ & $\ldots$ & XL26 \\
 &  &  &  & $\ldots$ & $\ldots$ & $748\pm37$ & $\ldots$ & KP26 \\
J092537.83+640921.7 & 141.40762 & 64.15603 & 0.052956 & $41.07$ & $185.52\pm3.15$ & $1101\pm48$ & $\ldots$ & JS26 \\
J103939.31+100253.1 & 159.91379 & 10.04808 & 0.161617 & $42.63$ & $413.75\pm4.32$ & $2047\pm187$ & $\ldots$ & JS26 \\
J112611.63+425246.4 & 171.54846 & 42.87956 & 0.156368 & $42.73$ & $497.29\pm4.75$ & $4501\pm51$ & $\ldots$ & JS26 \\
J153539.25+564406.4 (e0) & 233.91354 & 56.73511 & 0.207771 & $43.24$ & $436.14\pm3.25$ & $3627\pm50$ & $\ldots$ & JS26 \\
(e1) &  &  &  & $43.20$ & $472.18\pm2.42$ & $4270\pm37$ & $\ldots$ & JS26 \\
(e2) &  &  &  & $43.19$ & $485.49\pm2.98$ & $4005\pm45$ & $\ldots$ & JS26 \\
J154511.30+223856.1 (e0) & 236.29708 & 22.64892 & 0.218694 & $42.77$ & $358.03\pm3.89$ & $712\pm55$ & $\ldots$ & JS26 \\
(e1) &  &  &  & $42.73$ & $422.84\pm7.52$ & $660\pm23$ & $\ldots$ & JS26 \\
J222024.59+010931.3 (e0) & 335.10246 & 1.15869 & 0.212220 & $43.39$ & $367.47\pm5.15$ & $4941\pm48$ & $\ldots$ & JS26 \\
(e1) &  &  &  & $43.40$ & $436.48\pm2.59$ & $4324\pm31$ & $\ldots$ & JS26 \\
(e2) &  &  &  & $43.63$ & $505.17\pm4.57$ & $4155\pm45$ & $\ldots$ & JS26 \\
\enddata
\tablecomments{The table retains every reported study and epoch. Parenthetical ID entries distinguish epochs for multi-epoch objects. $\mathrm{FWHM}^{\mathrm{obs}}$ denotes the FWHM of the full fitted broad profile, after instrumental-resolution correction where reported. $\mathrm{FWHM}^{\mathrm{int}}$ is listed only when a study explicitly reports the underlying unscattered Gaussian in an electron-scattering model. Ellipses indicate quantities not reported by the cited study. Reference codes: this work = measurements presented here; IJ26 = \citet{Juodzbalis+2026}; JM26 = \citet{Matthee+2026}; KD26 = \citet{Davis+2026}; CC26 = \citet{Chen+2026}; AT25 = \citet{Taylor+2025}; JM24 = \citet{Matthee+2024}; IL24 = \citet{Labbe+2024}; AdG25 = \citet{deGraaff+2025b}; FD26a = \citet{DEugenioQSO1+2026}; FD26b = \citet{DEugenio+2026}; AT26a = \citet{Torralba+2026}; AT26b = \citet{TorralbaBHS+2026}; EL25 = \citet{LambridesAbs+2025}; FD25 = \citet{DEugenioIrony+2025}; IJ24 = \citet{Juodzbalis+2024}; XL25 = \citet{XLin+2025}; XJ26 = \citet{JiLRD+2026}; XL26 = \citet{XLin+2026}; KP26 = \citet{Park+2026}; LI26 = \citet{Ivey+2026}; JS26 = \citet{Shangguan+2026}.}
\end{deluxetable*}

\startlongtable
\begin{deluxetable*}{lcccccc}
\tabletypesize{\scriptsize}
\tablecaption{Reported H$\alpha$ Absorption Properties\label{tab:balmer_abs_properties}}
\tablewidth{0pt}
\tablehead{
\colhead{ID} & \colhead{$\mathrm{EW}_{\mathrm{H}\alpha,\mathrm{abs}}$} & \colhead{$\Delta v_{\mathrm{abs}}$} & \colhead{$\mathrm{FWHM}_{\mathrm{abs}}$} & \colhead{$\log\tau_{0,\mathrm{H}\alpha}$} & \colhead{$C_f$} & \colhead{Ref.} \\
\colhead{} & \colhead{\AA} & \colhead{km s$^{-1}$} & \colhead{km s$^{-1}$} & \colhead{} & \colhead{} & \colhead{}
}
\startdata
EGS-47962 & $8.4_{-5.1}^{+9.6}$ & $-204_{-71}^{+69}$ & $319_{-108}^{+116}$ & $0.76_{-0.36}^{+0.55}$ & $0.67_{-0.15}^{+0.16}$ & this work \\
JADES-GN-68797 & $5.3_{-1.9}^{+2.2}$ & $-228_{-9}^{+5}$ & $109_{-6}^{+31}$ & $1.84_{-0.87}^{+0.14}$ & $0.85_{-0.08}^{+0.05}$ & this work \\
 & $8.50_{-0.31}^{+0.27}$ & $-215_{-6}^{+6}$ & $438_{-7}^{+7}$ & $0.281_{-0.026}^{+0.005}$ & $0.984_{-0.022}^{+0.012}$ & IJ26 \\
 & $11.9$ & $-25$ & $\ldots$ & $\ldots$ & $\ldots$ & JM26 \\
 & $\ldots$ & $-368.72_{-7.33}^{+8.91}$ & $145_{-18}^{+33}$ & $>0.13$ & $0.64_{-0.02}^{+0.04}$ & CC26 \\
RUBIES-EGS-49140 & $6.5_{-3.1}^{+3.7}$ & $-29_{-12}^{+12}$ & $182_{-40}^{+34}$ & $0.83_{-0.32}^{+0.45}$ & $0.89_{-0.14}^{+0.08}$ & this work \\
 & $8.02_{-1.64}^{+2.19}$ & $3.7_{-12.0}^{+12.0}$ & $292_{-24}^{+31}$ & $0.788_{-0.157}^{+0.140}$ & $0.622_{-0.129}^{+0.192}$ & IJ26 \\
 & $11.1$ & $-22$ & $\ldots$ & $\ldots$ & $\ldots$ & JM26 \\
 & $5.1_{-0.7}^{+1.1}$ & $-14.0_{-6.4}^{+5.7}$ & $247_{-23}^{+28}$ & $\ldots$ & $\ldots$ & KD26 \\
 & $\ldots$ & $-22.23_{-14.69}^{+13.50}$ & $202_{-38}^{+32}$ & $>0.12$ & $>0.52$ & CC26 \\
 & $\ldots$ & $-50_{-15}^{+31}$ & $\ldots$ & $\ldots$ & $\ldots$ & AT25 \\
 & $341_{-20}^{+23}$ & $-55.90_{-4.02}^{+4.67}$ & $336_{-10}^{+8}$ & $\ldots$ & $\ldots$ & EL25 \\
(blue) & $9.3_{-0.4}^{+0.3}$ & $-46_{-5}^{+4}$ & $250\pm12$ & $0.45_{-0.05}^{+0.06}$ & $0.59\pm0.03$ & FD25 \\
(red) & $0.7_{-0.1}^{+0.2}$ & $160\pm10$ & $188\pm24$ & $-0.89_{-0.11}^{+0.14}$ & $0.86\pm0.08$ & FD25 \\
RUBIES-EGS-42046 & $7.2_{-1.7}^{+5.6}$ & $-375_{-17}^{+20}$ & $268_{-27}^{+129}$ & $0.39_{-0.08}^{+0.15}$ & $0.92_{-0.06}^{+0.05}$ & this work \\
 & $18.63_{-0.52}^{+0.49}$ & $-284_{-39}^{+32}$ & $619_{-35}^{+35}$ & $0.307_{-0.077}^{+0.101}$ & $0.928_{-0.064}^{+0.050}$ & IJ26 \\
 & $11.9$ & $-217$ & $\ldots$ & $\ldots$ & $\ldots$ & JM26 \\
 & $11.0_{-0.3}^{+0.7}$ & $-315.0_{-9.0}^{+58.0}$ & $612_{-16}^{+24}$ & $\ldots$ & $\ldots$ & KD26 \\
 & $\ldots$ & $-500.82_{-11.54}^{+12.52}$ & $240_{-19}^{+26}$ & $>0.35$ & $0.74_{-0.02}^{+0.02}$ & CC26 \\
 & $\ldots$ & $-184_{-14}^{+24}$ & $\ldots$ & $\ldots$ & $\ldots$ & AT25 \\
RUBIES-EGS-55604 & $7.0_{-2.1}^{+2.7}$ & $-257_{-6}^{+8}$ & $152_{-20}^{+30}$ & $1.18_{-0.40}^{+0.41}$ & $0.97_{-0.04}^{+0.02}$ & this work \\
 & $5.45_{-0.18}^{+0.18}$ & $-305_{-5}^{+4}$ & $174_{-12}^{+12}$ & $0.433_{-0.056}^{+0.050}$ & $0.794_{-0.028}^{+0.034}$ & IJ26 \\
 & $6.7$ & $-212$ & $\ldots$ & $\ldots$ & $\ldots$ & JM26 \\
 & $\ldots$ & $-268.36_{-34.94}^{+45.21}$ & $165_{-25}^{+27}$ & $>0.52$ & $0.85_{-0.03}^{+0.04}$ & CC26 \\
DeepDive-20910 & $4.7_{-2.9}^{+2.7}$ & $10_{-19}^{+85}$ & $113_{-46}^{+24}$ & $1.15_{-0.53}^{+0.53}$ & $0.88_{-0.14}^{+0.09}$ & this work \\
DeepDive-51909 & $4.7_{-1.3}^{+1.5}$ & $-91_{-9}^{+12}$ & $96_{-12}^{+14}$ & $1.34_{-0.39}^{+0.44}$ & $0.97_{-0.05}^{+0.02}$ & this work \\
GN-9771 & $8.8_{-0.2}^{+0.2}$ & $-21_{-1}^{+1}$ & $286_{-2}^{+2}$ & $0.46_{-0.02}^{+0.01}$ & $1.00_{-0.00}^{+0.00}$ & this work \\
 & $3.47_{-0.08}^{+0.08}$ & $-199_{-57}^{+57}$ & $162_{-38}^{+38}$ & $0.220_{-1.317}^{+0.290}$ & $0.466_{-0.510}^{+0.510}$ & IJ26 \\
 & $3.9$ & $-63$ & $\ldots$ & $\ldots$ & $\ldots$ & JM26 \\
 & $3.4\pm0.4$ & $-340$ & $240$--$280$ & $\ldots$ & $\ldots$ & JM24 \\
 & $\ldots$ & $-115\pm3$ & $220\pm5$ & $0.09_{-0.03}^{+0.03}$ & $\ldots$ & AT26a \\
BT-159717 & $6.5_{-1.7}^{+2.3}$ & $-27_{-3}^{+3}$ & $212_{-17}^{+13}$ & $0.72_{-0.15}^{+0.22}$ & $0.81_{-0.09}^{+0.10}$ & this work \\
 & $6.54_{-0.45}^{+0.39}$ & $-3.5_{-4.0}^{+4.0}$ & $257_{-16}^{+16}$ & $0.436_{-0.178}^{+0.126}$ & $0.828_{-0.125}^{+0.125}$ & IJ26 \\
 & $8.3_{-0.4}^{+0.5}$ & $-13_{-4}^{+5}$ & $254\pm19$ & $0.49_{-0.08}^{+0.09}$ & $0.95_{-0.06}^{+0.03}$ & FD26b \\
UNCOVER-4286 & $1.8_{-1.1}^{+1.3}$ & $-94_{-7}^{+9}$ & $45_{-10}^{+13}$ & $1.17_{-0.60}^{+0.55}$ & $0.86_{-0.21}^{+0.11}$ & this work \\
GN-15498 & $2.4_{-1.1}^{+1.5}$ & $-140_{-7}^{+9}$ & $56_{-11}^{+16}$ & $1.23_{-0.62}^{+0.56}$ & $0.89_{-0.07}^{+0.06}$ & this work \\
 & $6.10_{-0.25}^{+0.26}$ & $-47_{-5}^{+5}$ & $158_{-7}^{+7}$ & $0.775_{-0.090}^{+0.096}$ & $0.931_{-0.053}^{+0.044}$ & IJ26 \\
 & $6.5$ & $-36$ & $\ldots$ & $\ldots$ & $\ldots$ & JM26 \\
GS-13971 & $4.8_{-1.5}^{+1.7}$ & $-161_{-16}^{+26}$ & $169_{-21}^{+17}$ & $0.61_{-0.22}^{+0.19}$ & $0.80_{-0.04}^{+0.07}$ & this work \\
 & $6.31_{-0.23}^{+0.22}$ & $-146_{-36}^{+36}$ & $235_{-31}^{+31}$ & $0.340_{-0.126}^{+0.097}$ & $0.910_{-0.130}^{+0.130}$ & IJ26 \\
 & $5.7$ & $-81$ & $\ldots$ & $\ldots$ & $\ldots$ & JM26 \\
RUBIES-182791 & $4.7_{-2.0}^{+3.6}$ & $-216_{-32}^{+38}$ & $223_{-29}^{+35}$ & $0.34_{-0.18}^{+0.26}$ & $0.77_{-0.13}^{+0.15}$ & this work \\
 & $5.44_{-0.92}^{+1.07}$ & $-298_{-54}^{+48}$ & $306_{-57}^{+45}$ & $0.733_{-0.295}^{+0.190}$ & $0.420_{-0.050}^{+0.060}$ & IJ26 \\
 & $5.0$ & $-90$ & $\ldots$ & $\ldots$ & $\ldots$ & JM26 \\
BT-J1148 & $1.8_{-0.6}^{+1.0}$ & $100_{-7}^{+7}$ & $76_{-9}^{+17}$ & $0.32_{-0.09}^{+0.17}$ & $0.90_{-0.11}^{+0.07}$ & this work \\
 & $2.78_{-0.07}^{+0.08}$ & $57_{-2}^{+2}$ & $137_{-5}^{+5}$ & $0.628_{-0.056}^{+0.056}$ & $0.520_{-0.020}^{+0.020}$ & IJ26 \\
 & $3.6\pm0.8$ & $50$ & $240$--$280$ & $\ldots$ & $\ldots$ & JM24 \\
JADES-GN-28074 & $8.73_{-0.07}^{+0.07}$ & $-425_{-1}^{+1}$ & $483_{-2}^{+2}$ & $0.107_{-0.003}^{+0.007}$ & $0.994_{-0.008}^{+0.004}$ & IJ26 \\
 & $\ldots$ & $-321.30_{-4.05}^{+3.99}$ & $226_{-10}^{+15}$ & $>0.32$ & $0.94_{-0.01}^{+0.01}$ & CC26 \\
 & $\ldots$ & $-351_{-16}^{+14}$ & $297_{-19}^{+14}$ & $0.94_{-0.07}^{+0.07}$ & $0.998_{-0.004}^{+0.001}$ & IJ24 \\
DH-GS-5070 & $10.59_{-0.62}^{+0.44}$ & $-17_{-3}^{+3}$ & $252_{-7}^{+7}$ & $0.822_{-0.051}^{+0.062}$ & $0.930_{-0.060}^{+0.040}$ & IJ26 \\
JADES-GN-38147 & $3.72_{-0.63}^{+1.02}$ & $-358_{-23}^{+31}$ & $170_{-57}^{+64}$ & $0.352_{-0.291}^{+0.257}$ & $0.596_{-0.129}^{+0.290}$ & IJ26 \\
 & $4.3$ & $-156$ & $\ldots$ & $\ldots$ & $\ldots$ & JM26 \\
 & $\ldots$ & $-279.38_{-24.91}^{+26.24}$ & $108_{-26}^{+51}$ & $>0.24$ & $>0.58$ & CC26 \\
 & $\ldots$ & $61\pm4$ & $122\pm14$ & $0.45_{-0.04}^{+0.04}$ & $0.67_{-0.13}^{+0.17}$ & LI26 \\
The Cliff & $3.62_{-0.06}^{+0.07}$ & $70_{-1}^{+1}$ & $217_{-2}^{+2}$ & $0.049_{-0.016}^{+0.019}$ & $0.976_{-0.025}^{+0.017}$ & IJ26 \\
 & $5.1$ & $49$ & $\ldots$ & $\ldots$ & $\ldots$ & JM26 \\
 & $\ldots$ & $49_{-21}^{+26}$ & $133_{-92}^{+95}$ & $\ldots$ & $\ldots$ & AdG25 \\
A2744-QSO1 & $1.68_{-0.17}^{+0.17}$ & $-28_{-5}^{+5}$ & $167_{-16}^{+19}$ & $0.668_{-0.203}^{+0.170}$ & $0.268_{-0.028}^{+0.032}$ & IJ26 \\
 & $5_{-1}^{+2}$ & $-40\pm10$ & $259_{-24}^{+47}$ & $0.08_{-0.12}^{+0.18}$ & $0.56\pm0.07$ & FD26a \\
A2744-45924 & $8.0$ & $54$ & $\ldots$ & $\ldots$ & $\ldots$ & JM26 \\
(blue) & $1.2$--$2.0$ & $-143$ & $240$--$330$ & $\ldots$ & $\ldots$ & IL24 \\
(red) & $2.3$--$4.4$ & $172$ & $240$--$330$ & $\ldots$ & $\ldots$ & IL24 \\
OCEANS-102364 & $2.1_{-0.2}^{+0.3}$ & $-138.0_{-10.0}^{+13.0}$ & $151_{-14}^{+16}$ & $\ldots$ & $\ldots$ & KD26 \\
OCEANS-100424 & $1.10_{-0.14}^{+0.15}$ & $137.0_{-6.9}^{+7.3}$ & $122_{-13}^{+15}$ & $\ldots$ & $\ldots$ & KD26 \\
RUBIES-UDS-146995 & $\ldots$ & $-32_{-23}^{+22}$ & $\ldots$ & $\ldots$ & $\ldots$ & AT25 \\
RUBIES-EGS-28812 & $\ldots$ & $-172_{-32}^{+30}$ & $\ldots$ & $\ldots$ & $\ldots$ & AT25 \\
PAN-BH*-1 & $12.2\pm0.2$ & $-94\pm4$ & $283\pm8$ & $0.98_{-0.05}^{+0.07}$ & $1.00$ & AT26b \\
J102530.29+140207.3 (Egg) & $2.54\pm0.07$ & $-134$ & $138\pm5$ & $\ldots$ & $\ldots$ & XL25 \\
 & $\ldots$ & $-62\pm1$ & $188\pm5$ & $0.51_{-0.04}^{+0.04}$ & $0.86_{-0.02}^{+0.03}$ & XJ26 \\
J102208.52+084156.1 (blue) & $0.91\pm0.24$ & $-95$ & $94\pm19$ & $\ldots$ & $\ldots$ & XL25 \\
(red) & $2.04\pm0.60$ & $102$ & $227\pm45$ & $\ldots$ & $\ldots$ & XL25 \\
J012930.87+062843.32 & $2.8_{-0.3}^{+0.3}$ & $-66.6_{-5.9}^{+5.6}$ & $120_{-7}^{+7}$ & $\ldots$ & $1.00$ & XL26 \\
 & $\ldots$ & $-66\pm8$ & $\ldots$ & $0.61_{-0.20}^{+0.14}$ & $\ldots$ & KP26 \\
J082606.37-010001.31 & $6.4_{-0.5}^{+0.5}$ & $-78.4_{-8.7}^{+7.4}$ & $133_{-45}^{+35}$ & $\ldots$ & $0.91_{-0.06}^{+0.05}$ & XL26 \\
J082921.37+131237.44 & $6.2_{-2.1}^{+2.0}$ & $-21.7_{-15.0}^{+12.1}$ & $448_{-67}^{+50}$ & $\ldots$ & $1.00$ & XL26 \\
J094411.31-024908.65 & $4.2_{-1.7}^{+0.8}$ & $-277.7_{-21.6}^{+15.3}$ & $40_{-15}^{+25}$ & $\ldots$ & $0.38_{-0.06}^{+0.04}$ & XL26 \\
J102553.75+502843.24 & $5.2_{-0.2}^{+0.2}$ & $-237.2_{-8.7}^{+7.0}$ & $103_{-23}^{+40}$ & $\ldots$ & $0.83_{-0.04}^{+0.07}$ & XL26 \\
J132137.00-021417.04 & $1.3_{-0.1}^{+0.1}$ & $-127.0_{-0.2}^{+0.4}$ & $45_{-5}^{+5}$ & $\ldots$ & $1.00$ & XL26 \\
J142337.59+520216.05 & $4.9_{-0.3}^{+0.5}$ & $-237.6_{-12.2}^{+14.5}$ & $216_{-50}^{+22}$ & $\ldots$ & $0.84_{-0.11}^{+0.08}$ & XL26 \\
J161102.44+091728.60 & $10.0_{-4.1}^{+1.5}$ & $-190.9_{-47.7}^{+13.1}$ & $73_{-7}^{+92}$ & $\ldots$ & $0.93_{-0.03}^{+0.03}$ & XL26 \\
J164102.65+070806.47 & $3.0_{-0.9}^{+1.0}$ & $-111.6_{-26.6}^{+30.3}$ & $167_{-93}^{+68}$ & $\ldots$ & $0.62_{-0.15}^{+0.24}$ & XL26 \\
J164637.91+142648.62 & $10.4_{-1.4}^{+0.8}$ & $-239.5_{-11.6}^{+12.9}$ & $446_{-47}^{+23}$ & $\ldots$ & $1.00$ & XL26 \\
J165450.36+033741.74 & $7.0_{-1.0}^{+0.6}$ & $-213.3_{-14.1}^{+7.7}$ & $261_{-43}^{+27}$ & $\ldots$ & $1.00$ & XL26 \\
J171741.74+380752.47 & $4.4_{-0.5}^{+0.5}$ & $-212.9_{-9.5}^{+11.4}$ & $245_{-35}^{+32}$ & $\ldots$ & $1.00$ & XL26 \\
 & $\ldots$ & $-230\pm6$ & $\ldots$ & $0.51_{-0.03}^{+0.03}$ & $\ldots$ & KP26 \\
J212725.88-044808.92 & $3.4_{-0.6}^{+0.7}$ & $41.1_{-3.6}^{+6.9}$ & $107_{-37}^{+65}$ & $\ldots$ & $0.58_{-0.09}^{+0.14}$ & XL26 \\
J225535.58+154216.29 & $7.6_{-1.4}^{+1.8}$ & $27.5_{-25.4}^{+44.9}$ & $198_{-60}^{+60}$ & $\ldots$ & $0.80_{-0.06}^{+0.11}$ & XL26 \\
J111943.20+021911.32 & $2.3_{-0.7}^{+0.9}$ & $-261.0_{-33.1}^{+22.7}$ & $128_{-55}^{+83}$ & $\ldots$ & $0.41_{-0.10}^{+0.21}$ & XL26 \\
J113734.35+552028.16 & $3.0_{-0.8}^{+1.4}$ & $-76.7_{-22.0}^{+27.9}$ & $113_{-55}^{+42}$ & $\ldots$ & $1.00$ & XL26 \\
 & $\ldots$ & $-143\pm19$ & $\ldots$ & $0.62_{-0.48}^{+0.22}$ & $\ldots$ & KP26 \\
J134317.81+393418.07 & $1.8_{-0.2}^{+0.4}$ & $-147.9_{-3.6}^{+4.2}$ & $60_{-33}^{+32}$ & $\ldots$ & $0.64_{-0.05}^{+0.19}$ & XL26 \\
 & $\ldots$ & $-99\pm8$ & $\ldots$ & $0.72_{-0.05}^{+0.05}$ & $\ldots$ & KP26 \\
J190954.15+583112.37 & $7.2_{-1.9}^{+3.1}$ & $-24.4_{-10.1}^{+8.5}$ & $265_{-50}^{+63}$ & $\ldots$ & $1.00$ & XL26 \\
 & $\ldots$ & $-75\pm24$ & $\ldots$ & $0.58$ & $\ldots$ & KP26 \\
J092537.83+640921.7 & $2.29\pm0.26$ & $-145.6\pm13.2$ & $101\pm12$ & $>1.36$ & $0.40\pm0.02$ & JS26 \\
J103939.31+100253.1 & $4.57\pm0.23$ & $-705.2\pm11.7$ & $352\pm23$ & $>1.26$ & $0.23\pm0.01$ & JS26 \\
J112611.63+425246.4 & $3.05\pm0.17$ & $-223.3\pm11.4$ & $216\pm11$ & $>1.47$ & $0.24\pm0.01$ & JS26 \\
J153539.25+564406.4 (e0) & $1.54\pm0.07$ & $-845.2\pm3.9$ & $86\pm13$ & $0.88\pm0.15$ & $0.44\pm0.05$ & JS26 \\
(e1) & $2.04\pm0.08$ & $-842.2\pm3.7$ & $127\pm9$ & $0.81\pm0.07$ & $0.41\pm0.02$ & JS26 \\
(e2) & $1.66\pm0.07$ & $-852.1\pm3.8$ & $114\pm9$ & $0.92\pm0.09$ & $0.35\pm0.02$ & JS26 \\
J154511.30+223856.1 (e0) & $2.16\pm0.12$ & $-406.8\pm4.9$ & $100\pm10$ & $1.15\pm0.11$ & $0.47\pm0.03$ & JS26 \\
(e1) & $3.23\pm0.04$ & $-394.3\pm1.2$ & $116\pm2$ & $1.36\pm0.05$ & $0.56\pm0.01$ & JS26 \\
J222024.59+010931.3 (e0) & $2.52\pm0.13$ & $-373.9\pm6.3$ & $296\pm21$ & $-0.24\pm0.15$ & $>0.35$ & JS26 \\
(e1) & $3.04\pm0.20$ & $-385.9\pm9.5$ & $510\pm34$ & $-0.50\pm0.09$ & $>0.57$ & JS26 \\
(e1, red) & $0.29\pm0.05$ & $1548.0\pm16.7$ & $79\pm24$ & $>1.34$ & $0.05\pm0.01$ & JS26 \\
(e2) & $3.02\pm0.13$ & $-374.1\pm6.2$ & $487\pm21$ & $-0.45\pm0.08$ & $>0.53$ & JS26 \\
(e2, red) & $0.69\pm0.03$ & $1626.6\pm3.7$ & $56\pm12$ & $0.14\pm0.28$ & $>0.28$ & JS26 \\
\enddata
\tablecomments{The table retains every reported study, epoch, and resolved absorption component. Parenthetical entries distinguish epochs or components.  Gaussian velocity dispersions have been converted using $\mathrm{FWHM}=2\sqrt{2\ln2}\,\sigma$, and the Doppler parameters of \citet{XLin+2026} using $\mathrm{FWHM}=2\sqrt{\ln2}\,b$. A value of $C_f=1$ without uncertainty denotes a fixed or adopted full-covering model. Ellipses indicate quantities not reported by the cited study. Reference codes: this work = measurements presented here; IJ26 = \citet{Juodzbalis+2026}; JM26 = \citet{Matthee+2026}; KD26 = \citet{Davis+2026}; CC26 = \citet{Chen+2026}; AT25 = \citet{Taylor+2025}; JM24 = \citet{Matthee+2024}; IL24 = \citet{Labbe+2024}; AdG25 = \citet{deGraaff+2025b}; FD26a = \citet{DEugenioQSO1+2026}; FD26b = \citet{DEugenio+2026}; AT26a = \citet{Torralba+2026}; AT26b = \citet{TorralbaBHS+2026}; EL25 = \citet{LambridesAbs+2025}; FD25 = \citet{DEugenioIrony+2025}; IJ24 = \citet{Juodzbalis+2024}; XL25 = \citet{XLin+2025}; XJ26 = \citet{JiLRD+2026}; XL26 = \citet{XLin+2026}; KP26 = \citet{Park+2026}; LI26 = \citet{Ivey+2026}; JS26 = \citet{Shangguan+2026}.}
\end{deluxetable*}

\clearpage

\bibliography{ref}{}
\bibliographystyle{aasjournalv7}

\end{document}